\documentclass[onecolumn,superscriptaddress,
 amsmath,amssymb,
 aps,
 pre,
]{revtex4-1}
\usepackage{dcolumn}% Align table columns on decimal point
\usepackage{bm}% bold math
\usepackage{graphicx}
\usepackage{comment}
\usepackage[caption=false]{subfig}
\usepackage{color}

\newcommand{\Blue}{\textcolor{black}}

\newcommand{\bea}{\begin{eqnarray}}

\newcommand{\eea}{\end{eqnarray}}

\begin{document}

\preprint{}
\title{The distribution of the maximum of independent resetting Brownian motions}% Force line breaks with \\
%\author{Marco Biroli}
%\affiliation{LPTMS, CNRS, Univ.  Paris-Sud,  Universit\'e Paris-Saclay,  91405 Orsay,  France}
\author{Alexander K. Hartmann}
\affiliation{Institut f\"ur Physik, Universit\"at Oldenburg, 26111 Oldenburg, Germany}
\author{Satya N. Majumdar}
\affiliation{LPTMS, CNRS, Univ.  Paris-Sud,  Universit\'e Paris-Saclay,  91405 Orsay,  France}
\author{Gr\'egory Schehr}
\affiliation{Sorbonne Universit\'e, Laboratoire de Physique Th\'eorique et Hautes Energies, CNRS UMR 7589, 4 Place Jussieu, 75252 Paris Cedex 05, France}

%\date{\today}

\begin{abstract}

The probability distribution of the maximum $M_t$ of a single resetting 
Brownian motion (RBM) of duration $t$ and resetting rate $r$, properly 
centred and scaled, is known to converge to the standard Gumbel 
distribution of the classical extreme value theory. This Gumbel law 
describes the typical fluctuations of $M_t$ around its average $\sim \ln 
(r t)$ for large $t$ on a scale of $O(1)$. Here we compute the 
large-deviation tails of this distribution when $M_t = O(t)$ and show that the 
large-deviation function has a singularity where the second derivative 
is discontinuous, signaling a dynamical phase transition. Then we 
consider a collection of independent RBMs with initial (and resetting) 
positions uniformly distributed with a density $\rho$ over the negative 
half-line. We show that the fluctuations in the initial positions of the 
particles modify the distribution of $M_t$. The average over the initial 
conditions can be performed in two different ways, in analogy with 
disordered systems: (i) the annealed case where one averages over all possible 
initial conditions and (ii) the quenched case where one considers only the 
contributions coming from typical initial configurations. We show that 
in the annealed case, the limiting distribution of the maximum is 
characterized by a new scaling function, different from the Gumbel law 
but the large-deviation function remains the same as in the single 
particle case. \Blue{In the quenched case, the limiting 
(typical) distribution and the large deviation regime remain the same as in the single 
particle case. However, in the quenched case, an additional scaling regime emerges 
where $M_t = O(1)$ and the distribution of $M_t$ in this regime is described by a new nontrivial
rate function.} Our analytical results, both for the typical as 
well as for the large-deviation regime of $M_t$, are verified 
numerically with extremely high precision, down to $10^{-250}$ for the 
probability density of $M_t$.

\end{abstract}

%\keywords{Suggested keywords}%Use showkeys class option if keyword
                              %display desired
\maketitle

\section{Introduction}

\begin{figure}
\includegraphics[width = 0.7\linewidth]{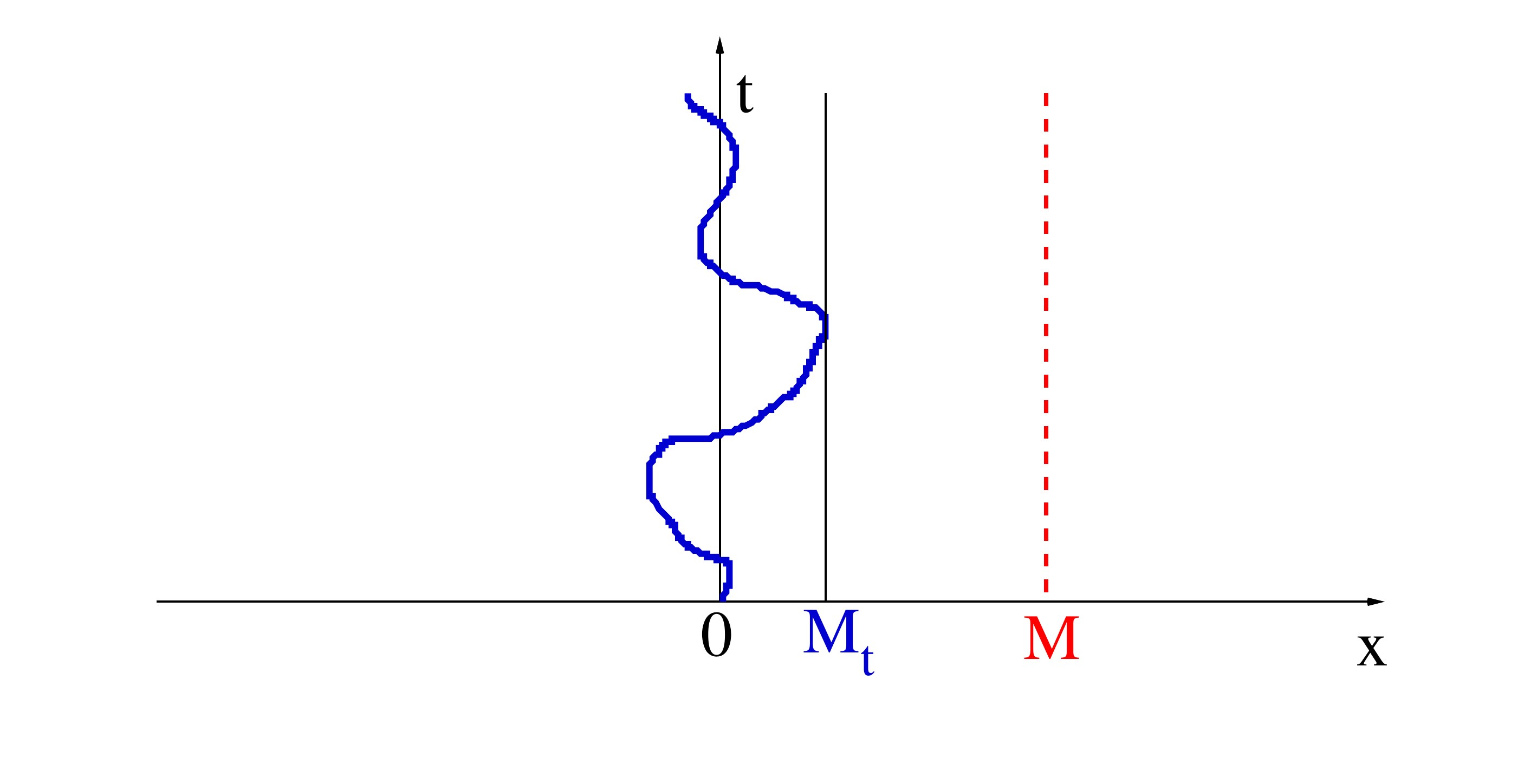}
\caption{Schematic trajectory of a stochastic search process $x(t)$, starting from $x(0)=0$, 
with global maximal value $M_t$ up to time $t$. A fixed target, shown by vertical dashed (red) line, is
located at $M$. The survival probability $Q(M,t)$ of the target up to time $t$ is exactly the probability of the event $M_t \leq M$, as in Eq.~(\ref{max.2}). }
\label{Fig_search}
\end{figure}

Consider a random searcher on a line whose dynamics
is characterized by a stochastic process $x(t)$, starting say at the origin $x=0$. Let
$M$ denote the location of a fixed target, see Fig.~\ref{Fig_search}. How long will the searcher take to find the
target? This is a key fundamental question for any search process. A quantity that
plays a central role is the survival probability $Q(M,t)$ of the target up to time $t$, i.e., the
probability that the target is not found by the searcher up to time $t$. The actual time 
to find the target is a random variable known as the 
first-passage time.
The probability distribution function (PDF) $P(t_f|M)$ of the first-passage time
is very simply related to the survival
probability by the relation
\bea
Q(M,t)= {\rm Prob.}\left[t_f\ge t\right] = \int_t^{\infty} P(t_f|M)\, dt_f\, ,
\label{surv.1}
\eea
which simply follows from the fact that if the target survives
up to $t$, it must be found by the searcher only after $t$, i.e., $t_f\ge t$. Taking
derivative with respect to $t$ relates the survival and the first-passage
probabilities
\bea
P(t_f|M)= -\frac{\partial}{\partial t} Q(M,t)\Big|_{t=t_f}\, .
\label{fp.1}
\eea
Search processes are ubiquitous in nature ~\cite{bell,viswanathan,berg}: 
animals searching for food, proteins searching for a DNA site
to bind, chemicals diffusing and searching for other constituents to react, 
debugging schemes searching for a bug in a computer program, randomized algorithms
searching for a global minimum in a high dimensional landscape and 
many others.
Survival probability and the associated first-passage time distribution
is a key concept in random search processes with
many applications~\cite{Condamin07,OLWB09,KMR10,benichou_11,MR15,GM16,GM16_b,GO17,GMO19,MB19}.
It can be considered as a cost function for a search process. The efficiency of a search process is
enhanced by minimizing the survival probability of the target. 
Given the ubiquity of search processes, the survival probability $Q(M,t)$ or the
associated first-passage probability $P(t_f|M)$ has been studied for more than
$100$ years across disciplines: in mathematics, physics, chemistry, biology and 
computer science~\cite{Feller_book,Redner_07,persistence_99,bf_05,Bray_13,Metzler_14}.
In physics of nonequilibrium many-body systems, the survival probability or `persistence'
also appears as a key quantity that characterizes the history dependence of the
underlying stochastic systems~\cite{Bray_13}.

Another, a priori unrelated, subject that has also been widely studied with applications ranging from sports,
climate science, finance all the way to physics of disordered systems is the extreme value
statistics (EVS)~\cite{Gum_58,Katz_02,Novak_11}, for a recent pedagogical review
on EVS, see Ref.~\cite{MPS_20}. In EVS, one typically looks at a time-series,
for example the temperature or rainfall data in a given weather station or 
the price of a stock, and records the
maximum (or minimum) $M_t$ of the process or the time-series up to time $t$. The histogram
of $M_t$ gives access to its PDF $P(M,t)$. The statistical behavior of $M_t$, encoded
in $P(M,t)$, gives important informations about the extreme fluctuations and play
a crucial role, e.g., in analysing data for global warming~\cite{RP06}. When the underlying
time-series is completely uncorrelated, the EVS is well understood from the classical
literature in statistics and mathematics~\cite{Gum_58,LLR_12} and $P(M,t)$, appropriately centered and scaled, is usually
governed by one of the three well known extreme value distributions, known as the
Gumbel, the Fr\'echet and the Wiebull class~\cite{MPS_20}.
However, much less is understood about
the EVS when the underlying time-series is correlated~\cite{MPS_20}.

These two subjects, namely the survival probability and the EVS, are actually very closely
related. To see this, consider again the one dimensional situation and let the process
$x(t)$ that represents the motion of the searcher in our underlying time-series, with initial
value $x(0)=0$. 
Let $M_t$ denote the maximum achieved by this process up to time $t$, i.e.,
\bea
M_t= \max\limits_{0\le \tau\le t}\left[\{x(\tau)\}\right]\, .
\label{max.1}
\eea
Then the cumulative distribution function (CDF) of the maximum is defined as
\bea
{\rm Prob.}[ M_t\le M] = \int_0^M P(M',t)\, dM' \, ,
\label{max_cdf.1}
\eea
where $P(M,t)$ is the PDF of $M_t$.
However, if the maximum $M_t$ is less than $M$, this event is equivalent to the probability
that all positions of the searcher $\{x(\tau)\}$ up to $t$ must be less than $M$ (see Fig.~\ref{Fig_search}). 
In other words,
this is precisely the probability that the target at $M$ is not touched/found by the process
up to $t$, i.e., the survival probability $Q(M,t)$ of the target up to time $t$. This
establishes a precise and general relation between the EVS and survival probability
\bea
{\rm Prob.}[ M_t\le M] = \int_0^M P(M',t)\, dM'= Q(M,t)\, .
\label{max.2}
\eea
Thus, if we know the survival probability $Q(M,t)$, the PDF of the maximum $M_t$ is simply
obtained from (\ref{max.2}) by taking a derivative with respect to $M$
\bea
P(M,t)= \frac{\partial}{\partial M} Q(M,t)\, .
\label{max.3}
\eea
To summarize, the survival probability $Q(M,t)$ has two arguments
$M$ and $t$. Taking a derivative with respect to $M$, as in Eq.~(\ref{max.3}),
provides the PDF of the maximum $M_t$ of the process up to time $t$.
In contrast, taking the negative of the derivative of $Q(M,t)$ with
respect to $t$, as in Eq.~(\ref{fp.1}), provides the PDF of the
first-passage time to the target at~$M$.

So far the discussion has been for any arbitrary random search process in
one dimension. We now focus on
a specific random search process that has created much interest over the last decade. \Blue{This} is known as
the resetting Brownian motion (RBM)~\cite{EM_11,EM_11b,EM_14}. In this simple model, the trajectory of
the searcher performs an ordinary Brownian motion, starting from its initial position at $0$, 
during a random time interval $\tau$, distributed via
$p(\tau)$. At the end of the interval, the trajectory is reset to
the starting position and a new random time interval is chosen from $p(\tau)$ during
which the searcher undergoes free diffusion and so on. 
When the interval between resettings
is distributed via  
\begin{equation}
p(\tau)=r\, e^{-r\, \tau}\,, \label{eq:poissonian}
\end{equation}
the model is called Poissonian resetting where the Brownian motion resets to its starting position at
a constant $r$~\cite{EM_11}. But other variants, such as `periodic' or `sharp' restart where $p(\tau)=\delta(\tau-T)$
or power-law distributed $p(\tau)$ have also been studied~\cite{NG16,PKE_16,BBR16,PR_17,ER20}. 

%\centerline{\rule{0.8\textwidth}{1pt}}

Here we will focus
on simple Poissonian resetting with rate $r$, mainly because this model
is simple and has a rather nice analytical structure.
A nonzero resetting rate $r$ has two major effects~\cite{EM_11}:
(i) by breaking the detailed balance, it drives the system into a nonequilibrium
stationary state where the position distribution is typically non-Gaussian at long times
and (ii) in the presence of a target at fixed $M$ and with the RBM starting and resetting at $0$, the
mean first-passage time (MFPT) to the target, as a function of the resetting rate $r$,
achieves a minimum at an optimal value $r^*$. The existence of an optimal $r^*$ makes
the RBM an efficient search process~\cite{EM_11,EM_11b,EM_14,KMSS_14,MSS_15,PKE_16,Reu_16,
MV_16,PR_17,CS_18,Bres_20,Pinsky_20,BMS_22} and confirms the scenario that when searching for a target in vain 
for a while, it is better to restart the search process: the rational being that the searcher
may explore a different pathway searching for the target, thus expediting the search process.
However, the existence of an optimal $r^*$ is not always guaranteed. Several models have
been studied, where by varying a parameter
in the underlying process, the optimal $r^*$ may undergo a phase transition, becoming zero
beyond a critical value of the 
parameter~\cite{KMSS_14,CS15,CM15,Reu_16,PR_17,Bel_18,RMS19,ANBBD19,Pal_19,Pal_19_2,Vasquez_20,Vasquez_22,
Vasquez_22_2,BMS23}.
Stochastic resetting has created much interest
across fields in the last decade (for recent reviews see Refs.~\cite{EMS_20,PKR_22,Gupta22}). Various
observables for diffusion with stochastic resetting have also been measured in experiments using holographic optical tweezers in one
and two dimensional optical traps~\cite{TPSRR_20,BBPMC_20,BFPCM_21,FBPCM_21}.

The purpose of this chapter is to illustrate the complementary perspectives of the same observable from
two different fields, i.e., search processes and the EVS, within the simple model of a resetting Brownian motion
as the search process. 
In Section II, we will explore the observable $Q_r(M,t)$ denoting the survival probability
of a target placed at $M$, in the presence of a single searcher whose dynamics 
is governed by the RBM, starting and resetting at
the origin with rate $r$. As mentioned earlier, $Q_r(M,t)$ can alternatively be interpreted as
the CDF of the maximum $M_t$ of the RBM (without any target) up to time $t$.
The Laplace transform (with respect to time $t$) of $Q_r(M,t)$ was already computed
exactly in Ref.~\cite{EM_11}. However, it turns out that the exact Laplace inversion is
difficult and interesting behaviors of $Q_r(M,t)$ for different scales of $M$ in real time $t$
are actually hidden in the Laplace transform. Treating $\partial_M Q_r(M,t)=P_r(M,t)$ as
the PDF of the maximum $M_t$ (see Eq.~(\ref{max.3})), one can extract the mean $\langle M_t\rangle$
and its variance explicitly and they have been used in many applications. These include, for instance, computing the statistics
of the number of distinct sites visited by a resetting random walker on a one dimensional 
lattice ~\cite{BMM22}, computing the mean perimeter and the mean area of the
convex hull of a two dimensional RBM~\cite{MMSS21}, and also calculating the
distribution of the time at which the maximum occurs in the stationary state of an 
RBM~\cite{MMS21,MMS22}. Similarly, higher moments of $M_t$ can also be extracted
analytically~\cite{SP21}.
However, extracting the full extreme-value distribution $Q_r(M,t)$ for all $M$ from its Laplace
transform is highly nontrivial.
Our first goal in Section II would be to analyze this Laplace transform
carefully to bring out the late time behavior of $Q_r(M,t)$ for all $M$. In particular, treating $Q_r(M,t)$
as the CDF of the maximum $M_t$ brings out new interesting scales of $M_t$. We will see that
while the typical fluctuations of $M_t$ are of $O(\ln t)$ for large $t$, there is an interesting
large-deviation tail when $M\sim O(t)$ where the associated rate function undergoes a second-order
dynamical phase transition. We will analyzanalyzee $Q_r(M,t)$ in different regions of the $M-t$ plane
and explore its scaling behaviors. 

In Section III, we will consider a generalization
of the single particle system to a many-particle scenario. Here we will have $N$ particles
distributed uniformly with density $\rho$ on the left half of the origin at $t=0$, with the
target located at $M\ge 0$. Each particle undergoes resetting Brownian motions
independently, i.e., the $i$-th particle starts at the initial random position $x_i$
and undergoes diffusion and resetting to its own initial position $x_i$.
The process ends when the target is found by any one of the $N$ particles.
Let $M_t(i)$ denote the maximum, up to $t$, of the $i$-th particle. Clearly,
$M_t(i)$'s are independent random variables.
Then the global maximum 
\bea
M_t= \max\limits_{1\le i\le N}\left[M_t(i)\right]\, ,
\label{max_global}
\eea
represents the maximum of a set of $N$ independent, but {\em not identically}
distributed variables. The different initial and resetting positions make the 
variables $M_t(i)$ non-identically distributed. We will see that this has interesting implications for 
the extreme-value statistics.
In addition, the behavior of the survival probability or the EVS depends strongly on how
the averaging over the initial positions is performed. If one averages the survival probability over all possible initial conditions with equal
weight, this is called {\em annealed} survival probability, in analogy with disordered systems. In contrast, when one averages only over
typical initial configurations (regular equidistant positions with typical interparticle separation $1/\rho$) and ignores the rest, the corresponding
survival probability is {\em quenched}, again in analogy with disordered systems.
We will see that the annealed and the quenched survival probabilities have rather different behaviors.
This also has interesting significance from the perspective of extreme-value statistics as will
be illustrated. Moreover, in the two cases, 
we will also explore both the typical as well as the large-deviation
regimes of the survival probability as a function of $M$ for large $t$. We note that
in Ref.~\cite{EM_11}, the survival probability for both the annealed and the quenched cases were
analyzed, but only for $M=0$, i.e., when the target is fixed at the origin. However, to understand
how the extreme $M_t$ behaves, we need to analyze the survival probability for general $M\ge 0$.
Thus, the results for general $M\ge 0$ presented in this chapter are original and go beyond Ref.~\cite{EM_11}.

In Section IV, we will verify our analytical predictions via high-precision numerical simulations.
Finally, we will conclude with some perspectives in Section V.

\section{Single particle case}

%\begin{figure}
%\includegraphics[width = 0.4\linewidth]{Fig_resetting.pdf}
%\caption{Schematic trajectory of a resetting Brownian motion with maximum $M_t$ at time $t$. The survival 
%probability $Q_r(M,t)$ is exactly the cumulative probability of the event $M_t \leq M$.}\label{Fig1}
%\end{figure}

We start with a single RBM, starting and resetting at the origin
with rate $r$, with a target fixed at $M$. Let $M_t$ be 
the maximum of this RBM up to time $t$, see Fig.~\ref{Fig_search}. As mentioned in the introduction,
the survival probability $Q_r(M,t)$ of the target up to time $t$ is 
exactly the CDF
of the random variable $M_t$, see Eq.~(\ref{max.2}). 
Thus $Q_r(M,t)$ has two alternative interpretations. When one thinks of it as the cumulative 
distribution of the maximum, one thinks $Q_r(M,t)$ as a function of $M$ for a fixed $t$. In contrast, when 
one thinks of $Q_r(M,t)$ as the survival probability of a fixed target located at $M$, one is more 
interested in determining how the survival probability decays as a function 
of time $t$, for fixed $M$. 
Here, we will therefore analyze $Q_r(M,t)$ in the $M-t$ plane for $M\geq 0$ and $t \geq 
0$. In the first case, we will fix $t$ and observe $Q_r(M,t)$ as a function of $M$ where it has the interpretation 
of a cumulative probability and increases from $0$ at $M=0$ to $1$ as $M \to \infty$, for any fixed $t$. 
In the second case, we will fix $M$ and study how $Q_r(M,t)$ decays as a function of time 
$t$.

\begin{figure}[t]
\includegraphics[angle=0,width= 0.6  \linewidth]{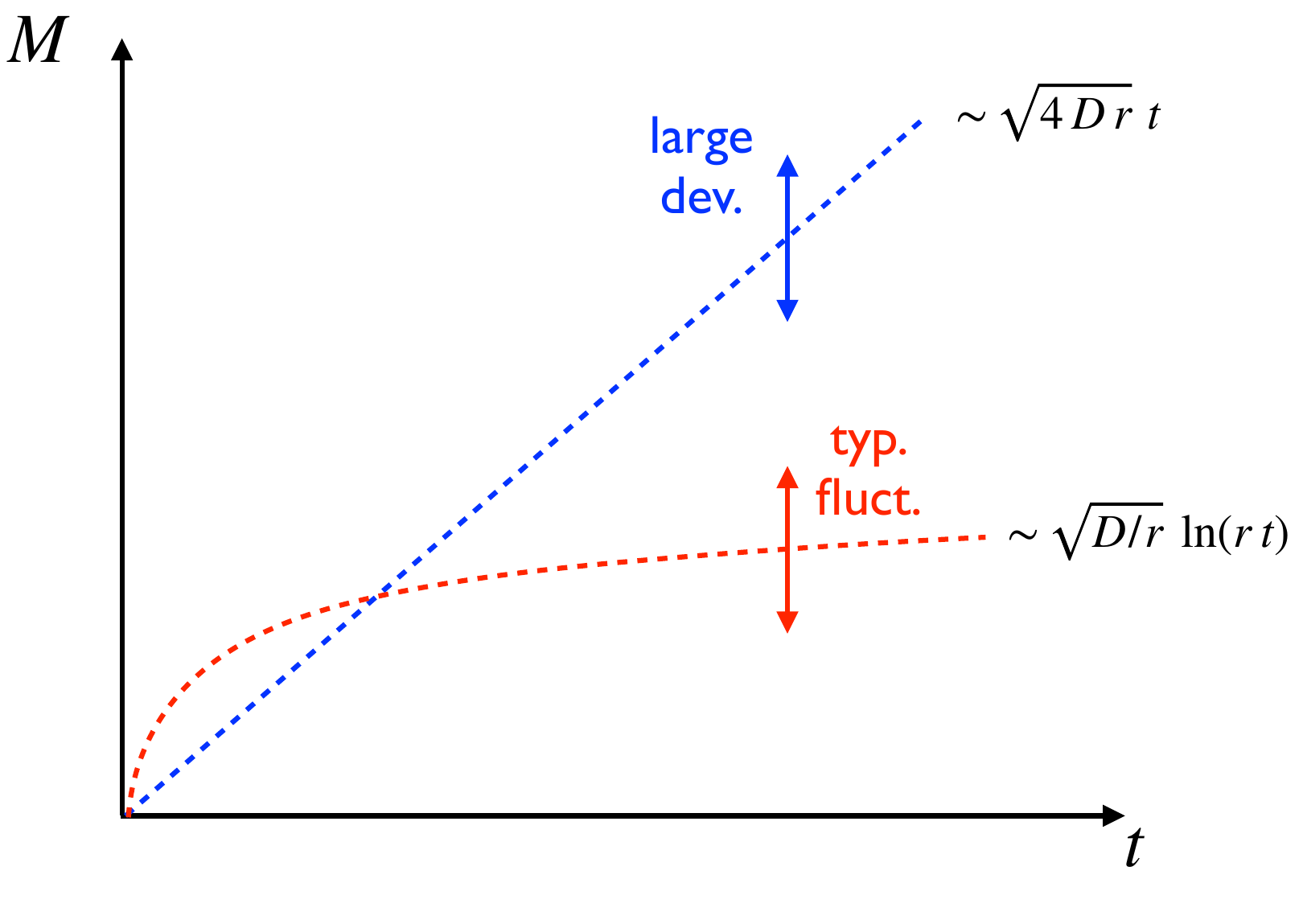}
\caption{A schematic representation of the two time scales associated with the fluctuations of the value of the maximum $M$ up to time $t$ for a resetting Brownian motion. For a fixed but large $t$, while the typical fluctuations scale as $\ln (rt)$ (shown by the red dashed curve), the large deviations scale linearly with $t$ as shown by the blue dashed line.}\label{FigMtPlane}
\end{figure}

The survival probability $Q_r(M,t)$, or instead its Laplace transform, can be easily computed using the 
renewal property of the RBM \cite{EM_11,EMS_20}. Let $Q_0(M,t)$ denote the survival probability of an ordinary 
Brownian motion without resetting. This Brownian motion starts at the origin and diffuses with a diffusion 
constant $D$ in the presence of a fixed target at $M \geq 0$. This survival probability can be easily 
computed by solving the diffusion equation with an absorbing boundary condition at $M$ and this well known 
result reads~\cite{Redner_07,bf_05,Bray_13}
\bea \label{survival_BM}
Q_0(M,t) = {\rm erf} \left( \frac{M}{\sqrt{4 D\,t}}\right) \;,
\eea
where ${\rm erf}(z)= (2/\sqrt{\pi})\int_0^x e^{-u^2}\,du$ is the error function. For large times, it 
decays as a power law, $Q_0(M,t) \sim {M}/{\sqrt{\pi D\,t}}$. In the presence of resetting with a nonzero 
rate $r>0$, one can write the survival probability $Q_r(M,t)$ in terms of $Q_0(M,t)$ via the renewal 
equation~\cite{EMS_20}
\bea \label{renewal}
Q_r(M,t) = e^{-r\,t} Q_0(M,t) + r\int_0^t d\tau  \, e^{-r \tau} Q_r(M,t-\tau) Q_0(M,\tau) \;.
\eea
This equation can be interpreted as follows. The first term corresponds to the event when there is no 
resetting in the full interval $[0,t]$ and the process is then an ordinary Brownian motion that needs to 
survive up to time $t$. The probability of this event 
equals the survival probability $Q_0(M,t)$ without resetting times
the probability that no resetting occurs in $[0,t]$. The probability
of no resetting in time $t$, using Eq.~(\ref{eq:poissonian}), 
is simply $\int_{t}^{\infty} p(\tau)d\tau=e^{-r\,t}$. This
then explains the first term
%the probability
%$\int_{t}^{\infty} p(\tau)d\tau=e^{-r\,t}$ that no resetting occurs up to time $t$,
% times the survival probability $Q_0(M,t)$, explaining the first term 
in Eq. (\ref{renewal}). The second term corresponds to the events with one
or more resettings. In this 
case let the last resetting event before $t$ occur at time $t-\tau$. 
In the interval $[t-\tau,t]$ there is 
no resetting, preceded by one resetting event at time $t-\tau$:
this joint event occurs with probability $p(\tau)d\tau=r\,d\tau 
e^{-r \tau}$. During this interval $[t-\tau,t]$, the process evolves as a free Brownian motion and its 
survival probability is simply $Q_0(M,\tau)$. But one needs to ensure also that the process (with 
resetting) did not cross $M$ during $[0,t-\tau]$, which occurs with a probability $Q_r(M,t-\tau)$. Since 
these events are independent due to the Markov nature of the process, we take the product of these three 
probabilities and integrate over $\tau$ from $0$ to $t$, leading to the second term in Eq.~
(\ref{renewal}). Let us now define the Laplace transform
\bea \label{Def_Lap_CDF}
\tilde Q_r(M,s) = \int_0^\infty e^{-st}\, Q_r(M,t) \, dt \;.
\eea 
Since Eq.~(\ref{renewal}) has a convolution structure in time, it is natural to take its Laplace transform 
which then leads to
\bea \label{renew_Lap}
\tilde Q_r(M,s) = \frac{\tilde Q_0(M,r+s)}{1 - r \tilde Q_0(M,r+s)} \;.
\eea
Using the expression of $Q_0(M,t)$ in Eq.~(\ref{survival_BM}), one has  
\bea \label{Def_Lap_CDFr0}
\tilde Q_0(M,s) = \int_0^\infty e^{-st}\, Q_0(M,t) \, dt = 
\frac{1}{s} \left(1 - e^{-\sqrt{s/D}\,M} \right) \;.
\eea 
Substituting this in Eq.~(\ref{renew_Lap}) gives the well known result~\cite{EM_11,EMS_20}
\bea \label{Lap_CDF}
\tilde Q_r(M,s) = \int_0^\infty e^{-st}\, Q_r(M,t) \, dt = 
\frac{1 - e^{-\sqrt{\frac{r+s}{D}}M}}{s + r\, e^{-\sqrt{\frac{r+s}{D}\,}M}} \;.
\eea 
Inverting this Laplace transform formally, we get
\bea \label{Brom_Qr}
Q_r(M,t) = \int_{\Gamma} \frac{ds}{2\pi i} \,e^{s\,t}\,  \frac{1 - e^{-\sqrt{\frac{r+s}{D}}M}}{s + r\, 
e^{-\sqrt{\frac{r+s}{D}\,}M}}
\eea
where $\Gamma$ is a Bromwich contour which has to pass to the right of all singularities of the integrand. 
In principle, this is an exact expression of the survival probability, or 
equivalently that of the CDF of $M_t$, 
valid for all $M$ and all $t$. In the rest of this section, we will analyze the behavior of $Q_r(M,t)$ in 
different regions of the $M-t$ plane, see Fig.~\ref{FigMtPlane}. In the following subsection, 
we analyze $Q_r(M,t)$ for fixed $t$ 
as a function of $M$ -- here the interpretation of $Q_r(M,t)$ is the cumulative distribution of the random 
variable $M_t$. In the next subsection, we will analyze $Q_r(M,t)$ as a function of $t$ for fixed $M$ -- 
this has the interpretation of the survival probability.

\subsection{Probability distribution of the maximum}

Here, we interpret $Q_r(M,t)$ as the CDF of the maximum $M_t$ up to time $t$ of the RBM, starting and 
resetting at the origin with rate $r$. It is also natural to consider the associated PDF $P_r(M,t) = 
\partial_M Q_r(M,t)$. Using Eq.~(\ref{Lap_CDF}), the Laplace transform of the PDF of 
$M_t$ is then given 
by
\bea \label{Lap_PDF}
\tilde P_r(M,s) = \frac{\partial}{\partial M} Q_r(M,s) = \frac{(s+r)^{3/2}}{\sqrt{D}} \frac{e^{-\sqrt{\frac{r+s}{D}\,}M}}{(s + r\, e^{-\sqrt{\frac{r+s}{D}\,}M})^2} \;.
\eea
Therefore $P_r(M,t)$ is given by the Bromwich integral
\bea \label{Br_PDF}
P_r(M,t) = \int_{\Gamma} \frac{ds}{2\pi i} \,e^{s\,t}\, \frac{(s+r)^{3/2}}{\sqrt{D}} \frac{e^{-\sqrt{\frac{r+s}{D}\,}M}}{(s + r\, e^{-\sqrt{\frac{r+s}{D}\,}M})^2} \;,
\eea
where $\Gamma$ is a Bromwich contour.

\vspace*{0.5cm}

We first fix $t$ to be large and consider $P_r(M,t)$ in Eq.~(\ref{Br_PDF}) as a function of $M$. To analyze 
this integral, we need to examine the singularities of the integrand in the complex $s$-plane and make 
sure that the vertical Bromwich contour $\Gamma$ is to the right of all the singularities. The integrand 
clearly has a branch cut at $s = -r$ and also a double pole at $s = s^*$ where $s^*$ is given by the root 
of the equation
\bea \label{sstar}
s + r\, e^{-\sqrt{\frac{r+s}{D}\,}M} = 0 \;.
\eea
In the limit of large $M$, we expect $|s^*| \ll 1$. Consequently we have, from Eq.~(\ref{sstar}), to 
leading order for large $M$,
\bea \label{s_star}
s^* = - r \, e^{-\sqrt{\frac{r}{D}}M} \;.
\eea
We also notice that the integrand in (\ref{Br_PDF}) admits, for large $t$ and large $M$, a saddle point 
$s_1$ which is obtained by minimizing $(s\,t - \sqrt{(r+s)/D}\,M)$, leading to
\bea \label{s1}
s_1 = -r + \frac{1}{4D} \left(\frac{M}{t}\right)^2 \;.
\eea
Therefore, when $s_1 > s^*$, we can deform the Bromwich contour to pass through $s_1$ and pick up the 
dominant contribution for large $t$ -- in this case we do not need to consider the contribution from the 
pole. In contrast, if $s_1 < s^*$, the dominant contribution will come from the pole at $s^*$. From 
(\ref{s_star}), we see that for large $M$, the pole $s^*$ is essentially close to zero, $s^* \approx 0$. 
Thus, from Eq.~(\ref{s1}), it follows that $s_1 > s^* \approx 0$ if $M > M_c(t)$ where
\bea \label{Mc}
M_c(t) = \sqrt{4\,D\,r}\,t \;.
\eea
In contrast, when $M<M_c(t)$, the leading contribution comes from the pole. By evaluating the residue at 
this pole, one gets
\bea \label{Gumbel.2}
P_r(M,t) &\approx& \sqrt{\frac{r}{D}}\,(r t)\, \exp\left[-\sqrt{\frac{r}{D}}\,M- r t \,e^{-\sqrt{\frac{r}{D}}M}\right]   \;.
\eea
Clearly, this is of order $O(1)$ when $M \sim \sqrt{D/r}\, \ln(r\,t)$. 

Hence we clearly see that there are two scales that characterise the fluctuations of $M_t$, namely when 
$M_t \sim \ln(r\,t)$ and $M_t \sim t$.  While the first one corresponds to the typical scale of 
fluctuations of $M_t$, the second one represents large deviations of $M_t$ compared to its typical value. 
We now consider these two scales of fluctuations separately.

\vspace*{0.5cm}

\noindent{\bf Typical fluctuations:} In this case, the dominant contribution to the integral in Eq.~
(\ref{Br_PDF}) comes from the double pole at $s = s^*$, leading to the asymptotic form of the PDF 
$P_r(M,t)$ in Eq.~(\ref{Gumbel.2}). This can be expressed in a nice scaling form
\bea \label{scaling_Gumbel}
P_r(M,t) \approx \sqrt{\frac{r}{D}} \, {\cal F}\left( \sqrt{\frac{r}{D}}  
\left( M - \sqrt{\frac{D}{r}}\, \ln(r\,t) \right) \right) \;,
\eea
where the scaling function ${\cal F}(v)$ has the Gumbel form
\bea \label{Gumbel_v}
{\cal F}(v)=e^{-v  -e^{-v}} \;,
\eea 
which is normalised to unity, $\int_{-\infty}^\infty {\cal F}(v)\, dv = 1$. From this scaling form 
(\ref{scaling_Gumbel}), it is clear that the PDF of $M_t$ is peaked around the typical value $M_t \sim 
\sqrt{\frac{D}{r}}\, \ln(r\,t)$ and the fluctuations around this typical value is of order $O(1)$ for 
large $t$. In other words, one can express the random variable $M_t$ as
\bea \label{Mt_v}
M_t = \sqrt{\frac{D}{r}} \left( \ln(r\,t) + v\right) \;,
\eea
where $v$ is a $O(1)$ random variable for large $t$, distributed via the Gumbel PDF ${\cal F}(v)$.

The result in Eq.~(\ref{Mt_v}) has a nice physical interpretation~\cite{EMS_20} in terms of extreme-value statistics (EVS) of 
independent and identically distributed (IID) random variables~\cite{MPS_20}. 
Consider the RBM without the target, starting
and resetting at the origin.
A typical trajectory of this RBM up to a large time $t$ consists of a large number of intervals separated by 
resetting events. Since the resetting occurs with rate $r$, the number of such intervals in time $t$ is typically 
$N_r = t/(1/r) = r\,t$. Within each interval labelled by $i = 1, 2, \cdots, N_r$, there is a local maximum of the 
displacement denoted by $m_i$. Clearly the global maximum $M_t$ is given by
\bea \label{max}
M_t = \max\{m_1, m_2, \cdots, m_{N_r} \} \;.
\eea
Since the $m_i$'s belong to different resetting intervals, they are clearly independent of each other.
To find the PDF of $m_i$, consider the $i$-th interval between resettings. Inside this interval,
the motion is purely Brownian and the CDF of the maximum within this interval of length say
$\tau$ is given by ${\rm erf}\left(\frac{m_i}{\sqrt{4D\tau}}\right)$ as in Eq.~(\ref{survival_BM}).
However, the interval duration $\tau$ itself is exponentially distributed via the PDF $p(\tau)= r\, e^{-r\tau}$.
Averaging over this random duration, one gets the CDF of the maximum within a given interval as
\bea
{\rm Prob.}\left[m_i\le m\right] = r\, \int_0^{\infty} e^{-r \tau}\, {\rm erf}\left(\frac{m}{\sqrt{4D\tau}}\right)\, d\tau=1-
e^{-\sqrt{r/D}\, m}\, ,
\label{block_max.cdf}
\eea
where we used Eq.~(\ref{Def_Lap_CDFr0}). Consequently, the PDF of $m_i$ is given by 
\bea
{\rm Prob.}\left[m_i=m\right]= \sqrt{\frac{r}{D}}\, e^{-\sqrt{r/D}\, m}\, ,  \quad\, m\ge 0\, .
\label{block_max_pdf} 
\eea
Hence, $M_t$ in Eq.~(\ref{max}) is the maximum of a set of $N_r=rt$ IID random variables, each distributed
via the PDF in Eq.~(\ref{block_max_pdf}).
Consequently, using the EVS of IID variables each with exponential distribution~\cite{MPS_20}, 
one would expect that $M_t$ will indeed scale as $\ln N_r = \ln 
(r\,t)$ with fluctuations of order $O(1)$ around it and in addition, the centered and scaled distribution will have the Gumbel 
from as in Eq.~(\ref{Gumbel_v}).

\vspace*{0.5cm}

\noindent{\bf Large-deviation regime:} Here we are interested in the large-deviation tail of $P_r(M,t)$ 
when $M = {\cal O}(t)$. In this limit, the saddle point $s_1$ is of order $O(1)$ from Eq.~ (\ref{s1}). If 
$M>M_c(t) = \sqrt{4D\,r}\,t$ given in Eq.~(\ref{Mc}), the saddle point $s_1 > s^* \approx 0$. In this 
case, deforming the Bromwich contour through the saddle point, and performing the steepest descent 
analysis, we get
\bea \label{Pr_largedev}
P_r(M,t) \approx e^{-r\,t - \frac{M^2}{4\,D\,t}} \;.
\eea 
Thus so far, we have seen that there are two well separated scales of fluctuations of $M_t$, respectively 
of order $O(t)$ and order $O(\ln(r\,t))$, where the PDF takes very different forms given respectively in 
Eqs. (\ref{Pr_largedev}) and (\ref{Gumbel.2}). One may wonder what happens in the intermediate regime when 
$\ln{t}\ll M \ll t$. In this regime, the Gumbel form in Eq.~(\ref{Gumbel.2}) continues to hold but one can 
neglect the term $r t \,e^{-\sqrt{r/D}\,M}$ since $M \gg \ln(r\,t)$. Hence the approximate form of the PDF 
in this intermediate range reads
\bea \label{Gumbel.3}
P_r(M,t) \approx \frac{r^{3/2}}{\sqrt{D}}\, t\, e^{-\sqrt{\frac{r}{D}}M}  \;.
\eea

\begin{figure}[t]
\includegraphics[angle=0,width= 0.6  \linewidth]{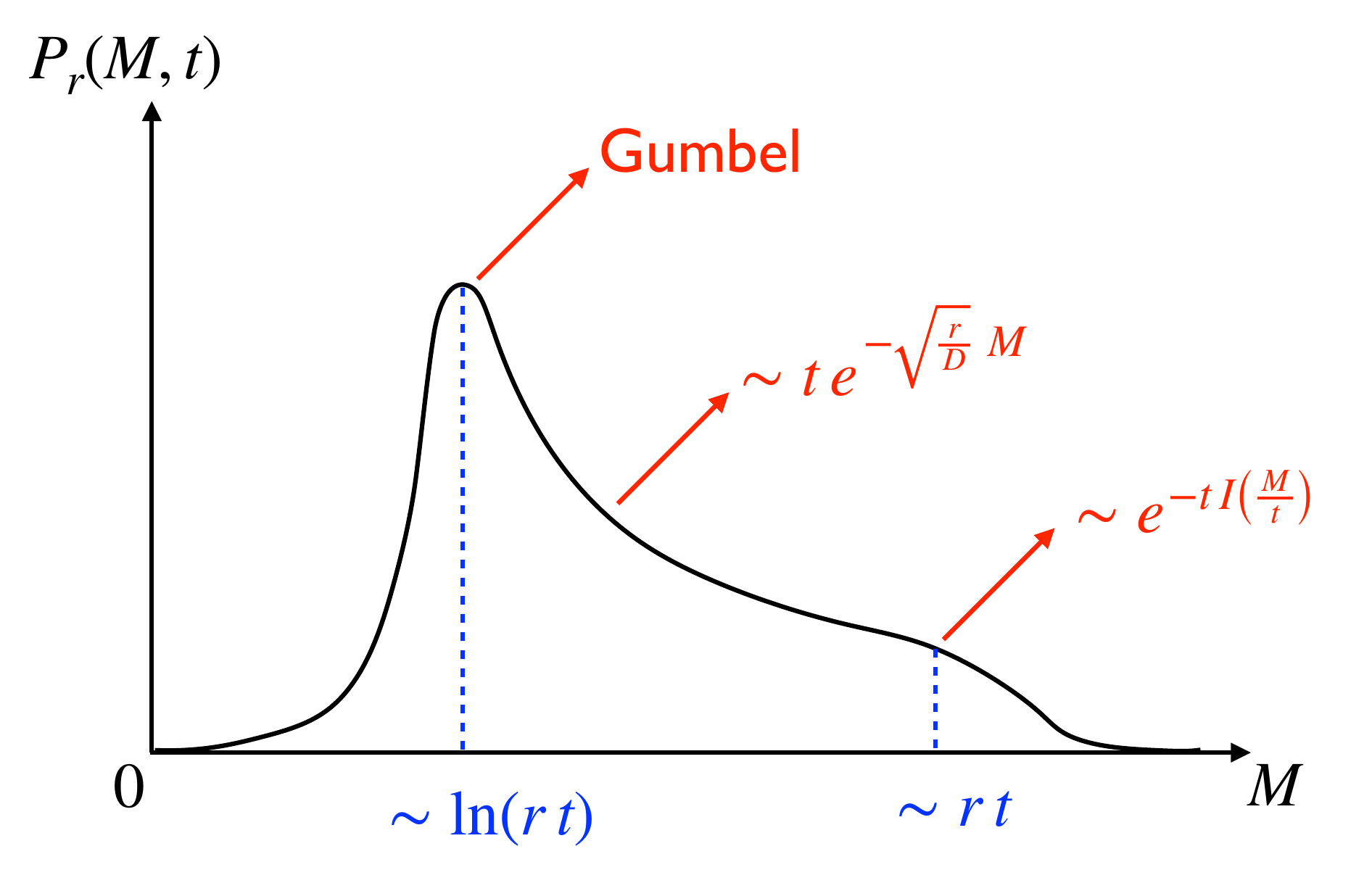}
\caption{A schematic plot of the PDF $P_r(M,t)$ of the maximum $M_t$ as a function of $M$, for fixed but large $t$ as in Eqs. \eqref{summary_PDF} and \eqref{2_3}. The two vertical dashed blue lines represent the typical and the large-deviation scales.}
\label{Fig_single_particle}
\end{figure} 
To summarise, the PDF of the maximum exhibits different behaviors in the three different regimes
\bea \label{summary_PDF}
P_r(M,t) \approx
\begin{cases}
& \sqrt{\frac{r}{D}}\, \exp{\left[-\sqrt{\frac{r}{D}}\left(M - \sqrt{\frac{D}{r}}\ln (r\,t)\right)-e^{-\sqrt{\frac{r}{D}}\left(M - \sqrt{\frac{D}{r}}\ln (r\,t)\right)}\right]} 
\quad {\rm for} \quad M = {\cal O}\left(\ln (r\,t)\right) \\
& \\
& \frac{r^{3/2}}{\sqrt{D}} \, t\, e^{-\sqrt{\frac{r}{D}}M} \quad, \quad {\rm for} \quad \ln{r\,t} \ll M \ll r\,t \\
& \\
& e^{-r\,t - \frac{M^2}{4\,D\,t}} \quad, \quad {\rm for} \quad M = {\cal O}(r\,t) \\
\end{cases}
\eea
While the Gumbel behavior in the first line corresponds to the typical trajectories with a large number 
of resettings, the last line corresponds to very rare trajectories that have undergone no reset up to time 
$t$. The intermediate regime corresponds to trajectories with few resets. The right tail of the 
intermediate regime and the left tail of the last regime can be combined into a single large-deviation 
form
\bea \label{2_3}
P_r(M,t) \approx  e^{- t \, I\left( \frac{M}{t}\right)} \;,
\eea  
where
\bea \label{rate_function}
I(z) = 
\begin{cases}
& \sqrt{\frac{r}{D}}\,z \quad, \quad z \leq z_c = \sqrt{4\,D\,r} \\
&\\
& r + \frac{z^2}{4D} \quad, \quad z \geq z_c = \sqrt{4\,D\,r} \;.
\end{cases}
\eea
Thus, the second derivative of $I(z)$ is discontinuous at $z = z_c$. In terms of the location of the 
maximum at time $t$, one sees that there is a ``light cone'' $M_c(t) = z_c\,t$ in the $M-t$ plane, such 
that for $M \leq M_c(t)$, the typical behavior of the maximum is established, while outside the ``light 
cone'' $M \geq M_c(t)$, the distribution of $M$ is still atypical, corresponding to rare trajectories with 
no resetting. Indeed, this large-deviation form of $P_r(M,t)$ in Eq.~(\ref{rate_function}) is exactly 
identical to the large-deviation form of the position distribution of the RBM~\cite{MSS_15}.
The typical, intermediate and the large-deviation behaviors of 
$P_r(M,t)$ are summarized schematically
in Fig.~\ref{Fig_single_particle}.

\subsection{Survival probability}

In this subsection, we consider $Q_r(M,t)$ as a survival probability, with a target fixed at $M$ and 
analyze it as a function of time $t$. This means traversing the $M-t$ plane in Fig.~\ref{FigMtPlane} 
horizontally by increasing $t$ at a fixed $M$.

We first consider the limit of large $t$, with $M$ fixed but not necessarily large. In this case, 
the dominant behvaior of the integral in Eq.~(\ref{Brom_Qr}) emerges 
from the pole of the integrand at $s= s^*$, where we recall from Eq.~(\ref{sstar}) that 
$s^*$ satisfies
\bea 
\label{sstar.2}
s^*  + r\,e^{-\sqrt{\frac{r+s^*}{D}}\,M} = 0 \;.
\eea
Evaluating the residue at the pole, we see that, up to pre-exponential factors, $Q_r(M,t)$ decays 
exponentially for large $t$ as
\bea \label{Q_larget}
Q_r(M,t) \sim e^{-\theta_r(M)\,t}  \quad, \quad {\rm where} \quad \theta_r(M) = - s^* \;.
\eea
It is easy to evaluate $\theta_r(M)$ as a function of $M$ by solving Eq.~(\ref{sstar.2}) numerically.
%as  shown in Fig.~(xxx). 
The asymptotic behaviors of $\theta_r(M)$ for small and large $M$ are given by
\bea \label{theta_of_M}
\theta_r(M) \approx
\begin{cases}
&r - \frac{r^2 M^2}{D} \quad, \quad\, M \to 0 \\
&\\
&r\,e^{-\sqrt{\frac{r}{D}}\,M} \quad, \quad M \to \infty \;.
\end{cases}
\eea 
As the target recedes, i.e., $M$ becomes large, the exponent $\theta_r(M)$ becomes exponentially small, 
indicating that the target survives longer, as expected. In this large $M$ regime, Eq.~(\ref{Q_larget}) 
can be written in the form
\bea \label{Gumbel_surv}
Q_r(M,t) \approx \exp{\left[- r t \,e^{-\sqrt{\frac{r}{D}}\,M}\right]} = 
\exp{\left[-\,e^{-\sqrt{\frac{r}{D}}\left(M- \sqrt{\frac{D}{r}}\,\ln(r\,t) \right)}\right]} \;.
\eea
This is exactly the CDF of the Gumbel distribution in Eq.~(\ref{Gumbel.2}), associated with the typical 
fluctuations of the random variable $M_t$.

In the opposite limit $M \to 0$, the exponent $\theta_r(M)$ in Eq.~(\ref{theta_of_M}) approaches a 
non-zero constant $r$. A priori, this may look a bit strange since one expects that $Q_r(M,t) \to 0$ as $M 
\to 0$ for any time $t$. Indeed, in Eq.~(\ref{Q_larget}), we have ignored pre-exponential factors. Taking 
into account these factors, it is easy to show that, as $M \to 0$,
\bea \label{Mto0}
Q_r(M,t) \approx e^{-r\,t}\frac{M}{\sqrt{\pi D\,t}} \;.
\eea  
Thus $Q_r(M,t)$ indeed vanishes as $M \to 0$, but the exponent $\theta_r(M=0)=r$ remains finite. Physically, 
this corresponds to Brownian trajectories that have not reset up to time $t$ and have escaped in the 
direction opposite to the target. This is because if the trajectory resets even once, the target at $M=0$ will not 
survive, since it is close to the resetting position at $0$.

So far we have discussed the large $t$ behavior of the survival probability $Q_r(M,t)$ for fixed $M$ and 
we have seen that when $t = {\cal O}(e^{\sqrt{\frac{r}{D}}M})$, 
it acquires the form of the CDF of a 
Gumbel variable in Eq.~(\ref{Gumbel_surv}). This corresponds to crossing the curve $M = \sqrt{D/r}\, \ln(r\,t)$ 
in the $M-t$ plane in Fig.~\ref{FigMtPlane}. One can also analyze the behavior of $Q_r(M,t)$ for 
intermediate values of $t$, e.g., when $t = O(M)$. In this case, we use the large-deviation behavior of 
the PDF $P_r(M,t)$ in Eq.~(\ref{2_3}). Integrating over $M$ gives
\bea \label{2_3_Pr}
Q_r(M,t) \approx 1 - B\, e^{-t\, I\left( \frac{M}{t}\right)} \;,
\eea
where $B$ is an unimportant pre-exponential factor (subleading for large $t$) and the rate function $I(z)$ 
is given in Eq.~(\ref{rate_function}). When $t<M/\sqrt{4D\,r}$, i.e., $z =M/t > z_c=\sqrt{4Dr}$, one gets using Eq.~
(\ref{rate_function})
\bea \label{Q_short_t}
Q_r(M,t) \approx 1 - B\,e^{-r\,t - \frac{M^2}{4D\,t}} \quad, \quad t<M/\sqrt{4D\,r} \;.
\eea
In contrast, when $t>M/\sqrt{4D\,r}$, but still $t \ll e^{\sqrt{\frac{r}{D}}M}$, using the first line of 
(\ref{rate_function}), one gets
\bea \label{Q_inter}
Q_r(M,t) \approx 1 - B\,e^{- \sqrt{\frac{r}{D}}\,M} \;.
\eea
Thus in this regime the survival probability depends only weakly on time through the prefactor $B$. 
Finally, when $t$ approaches $e^{\sqrt{\frac{r}{D}}M}$, it crosses over to 
the Gumbel form 
(\ref{Gumbel_surv}). To summarize, for fixed $M$ and as a function of $t$, there are two time scales
$t\sim O(M)$ and $t\sim e^{\sqrt{\frac{r}{D}}M}$, corresponding to crossing horizontally the two lines
in the $M-t$ plane in Fig.~\ref{FigMtPlane}: the first time scale $t\sim O(M)$ corresponds
to the regime of the large deviation of the EVS, while the second one corresponds to the
regime of the typical behavior of the EVS.

%\vspace*{2cm}

\section{Multi-particle case}

\begin{figure}
\includegraphics[width = 0.7\linewidth]{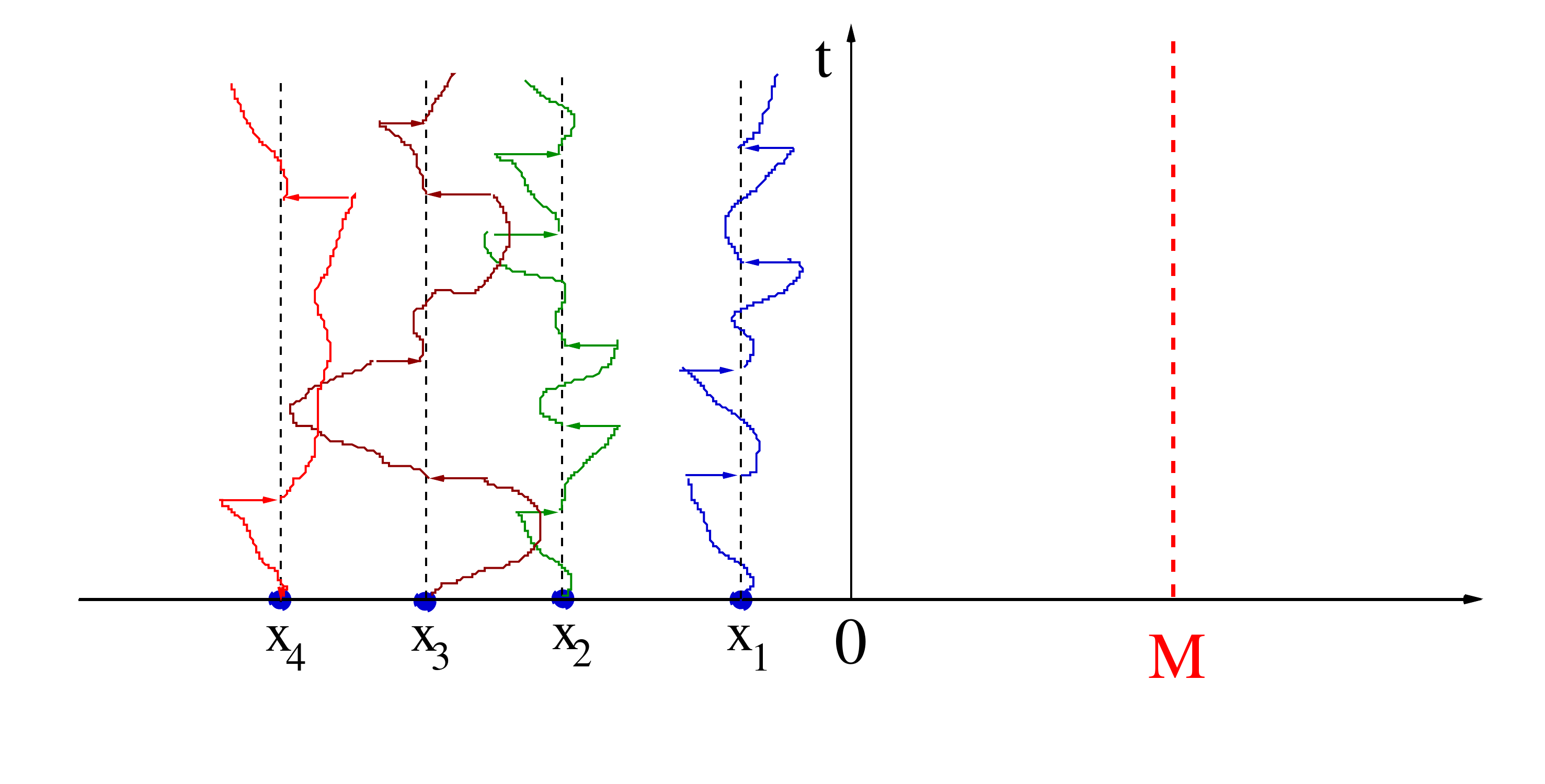}
\caption{Schematic trajectories of $4$ independent resetting Brownian motions up to time $t$, each starting and resetting at $x_i \leq 0$
($i=1$, $2$, $3$ and $4$). The target is located at $M>0$ shown by the dashed red vertical line.}
\label{Fig_traject} 
\end{figure}

We now consider $N$ independent RBMs, located initially on the negative
half line with initial positions $\{x_i\}$ ($i=1,2,\ldots, N$).
The $i$-th particle starts at $x_i$, undergoes independent diffusion
and resets to $x_i$ with rate $r$.
The initial position $x_i$ of 
the $i$-th particle is chosen independently from a uniform distribution over $x_i \in [-L,0]$. We are 
interested in the thermodynamic limit when $N \to \infty$, $L \to \infty$, with the ratio $N/L = \rho$ 
fixed. For a fixed set of initial positions, we define $M_t$ as the position of the global maximum of all $N$ 
particles up to time $t$. Let us denote the CDF of $M_t$ by $Q_r(M,t|\left\{ x_i\right\}) = {\rm Prob.}(M_t 
\leq M)$ for a given set of initial positions $\{ x_i\}$'s. This CDF is also the survival probability of a 
fixed target located at $M$, see Fig.~\ref{Fig_traject}. Here, we can interpret this search process as 
a team-search by $N$ independent walkers. The search terminates when any one of the $N$ walkers finds the 
target for the first time.

Using the independence of $N$ particles, the CDF for a fixed set of initial positions $\{x_i\}$'s is given by
\bea \label{Qr_N}
Q_r(M,t| \left\{x_i \right\}) = \prod_{i=1}^N  Q_r(M-x_i,t) \;,
\eea
where $Q_r(M,t)$ is the single particle CDF discussed in the previous section. 
Now we also need to average the survival probability in Eq.~(\ref{Qr_N}) with respect to the initial positions 
$\{ x_i\}$'s, drawn independently from a uniform distribution. This averaging can be done in two different ways, 
namely (i) annealed and (ii) quenched, akin to disordered systems. In fact, 
this setup has been used to 
compute the current distribution in nonequilibrium systems of independent stochastic processes, such as 
for diffusive particles~\cite{DG09,BMRS20,Marbach21,BJC22,Costan23,DMS23}, 
for run-and-tumble particles~\cite{BMRS20,BJC22,JRR23} and 
also for resetting Brownian motions~\cite{Costan23}. More recently,
the total number of visits to the origin by all particles up to time $t$ (equivalently the local time at the
origin for the combined $N$-particle process) has been computed
for the similar step-like initial condition~\cite{BMR23}.
Here, we will use the formalism 
developed above and adapt it to study the survival probability, or equivalently the CDF of 
the maximum $M_t$. In the annealed case, one averages $Q_r(M,t| \left\{x_i \right\})$ in Eq.~(\ref{Qr_N}) 
directly, i.e.,
\bea
Q_{\rm an}(M,t)= \overline{Q_r(M,t|\{x_i\})}= 
\left[Q_{\rm avg}(M,t)\right]^N\, ,
\label{an_avg.1}
\eea
where
\bea
Q_{\rm avg}(M,t)= \lim\limits_{L\to \infty} \frac{1}{L}\, \int_{-L}^0 
Q_r(M-x,t)\, dx
\label{an_avg.2}
\eea
%\remarkAKH{Here $Q_{\rm an}$ denoted the averaged single-particle cumulative
%distribution while below $Q_{\rm an}$ is the annealed multi-particle
%cumulative distribution. I use below $Q_{\rm avg}$ for the former....}

where $\overline{\cdots}$ denotes the averaging over the initial positions $\{x_i\}$'s.
In contrast, in the quenched case, one is interested in the typical initial configurations that dominate the 
survival probability. One way to extract the contribution from this typical initial configuration is to take the 
average over the logarithm
of $Q_r(M,t|\{x_i\})$ and then re-exponentiate, i.e.,
\bea
Q_{\rm qu}(M,t)= \exp\left[ \overline{\ln Q_r(M,t|\{x_i\})}\right]\, .
\label{qu_avg.1}
\eea
It turns out that this is equivalent to choosing a specific initial 
configuration with equispaced points separated by $1/\rho$~\cite{DG09,BMRS20}.

In the two sub-sections below, we discuss the annealed and the quenched cases separately. 

\vspace*{0.5cm}

\subsection{Annealed case}

In the annealed case, the survival probability or equivalently, the CDF $Q_{\rm an}(M,t)$ is given by
Eq.~(\ref{an_avg.1}), where 
the average $\overline{ \cdots} 
$ is over the initial positions $\{x_i \}$'s, each drawn independently and uniformly from $x_i \in 
[-L,0]$. We first replace $Q_r(M-x_i,t)$ by $1 - (1-Q_r(M-x_i,t))$ in 
Eq.~(\ref{Qr_N}) and average over $\{ 
x_i\}$'s as in Eq.~(\ref{an_avg.1}). This gives, using the
independence of $x_i$'s,
\bea \label{CFD_N.2}
Q_{\rm an}(M,t) =   \left( 1 - \frac{1}{L}\int_{-L}^0 (1-Q_r(M-x,t))\,dx \right)^N \;. 
\eea
Changing varible $x \to -x_0$ and taking the thermodynamic limit $N \to \infty$, $L \to \infty$, with the 
ratio $N/L = \rho$ fixed, we get
\bea \label{CDF_N_thermo}
Q_{\rm an}(M,t) = \exp{\left[- \rho\,\int_0^\infty \left(1 - Q_r(M+x_0,t)\right)\,dx_0\right]} \;.
\eea
We remark that the special case $M=0$, i.e., the survival probability $Q_{\rm an}(M=0,t)$ for
a target fixed at the origin was analyzed in Ref.~\cite{EM_11} with the interesting result that
at long times, $Q_r(M=0,t)$ decays algebraically as 
\bea
Q_{\rm an}(M=0,t) \sim t^{-\theta}\, , \quad \theta= \rho\, \sqrt{\frac{D}{r}}\, ,
\label{Qanm0.1}
\eea
with a continuously varying exponent $\theta$. Here, our goal is to go beyond $M=0$, and study
the full CDF $Q_r(M,t)$ for all $M$.

To proceed for general $M\ge 0$, we rewrite Eq.~(\ref{CDF_N_thermo}) as
\bea \label{CDF_N_thermo.2}
Q_{\rm an}(M,t) = e^{- \rho J(M,t)} \quad, \quad {\rm where}\quad 
J(M,t) = \int_M^\infty \left(1 - Q_r(x_0,t)\right)\,dx_0 \;.
\eea
We then define the Laplace transform of $J(M,t)$ with respect to $t$ as
\bea \label{Lap_J}
\tilde J(M,s) = \int_0^\infty e^{-st}\, J(M,t) \, dt \;.
\eea
Taking the Laplace transform of $J(M,t)$ in Eq.~(\ref{CDF_N_thermo.2}) and using the single particle 
result in Eq.~(\ref{Lap_CDF}), we get
\bea \label{Lap_J2}
\tilde J(M,s) = \frac{\sqrt{D(r+s)}}{r\,s} \ln \left(1 + \frac{r}{s}\, e^{-\sqrt{\frac{r+s}{D}}M} \right) \;.
\eea
Using Bromwich inversion formula, we can write $J(M,t)$ formally as
\bea \label{J_Mt}
J(M,t) = \int_\Gamma \frac{ds}{2\pi i}\, e^{st} \, \frac{\sqrt{D(r+s)}}{r\,s} \ln \left(1 + \frac{r}{s}\, e^{-\sqrt{\frac{r+s}{D}}M} \right) \;,
\eea
where $\Gamma$ is the Bromwich contour.

\vspace*{0.5cm}

\begin{figure}[t]
\includegraphics[angle=-0,width = 0.6\linewidth]{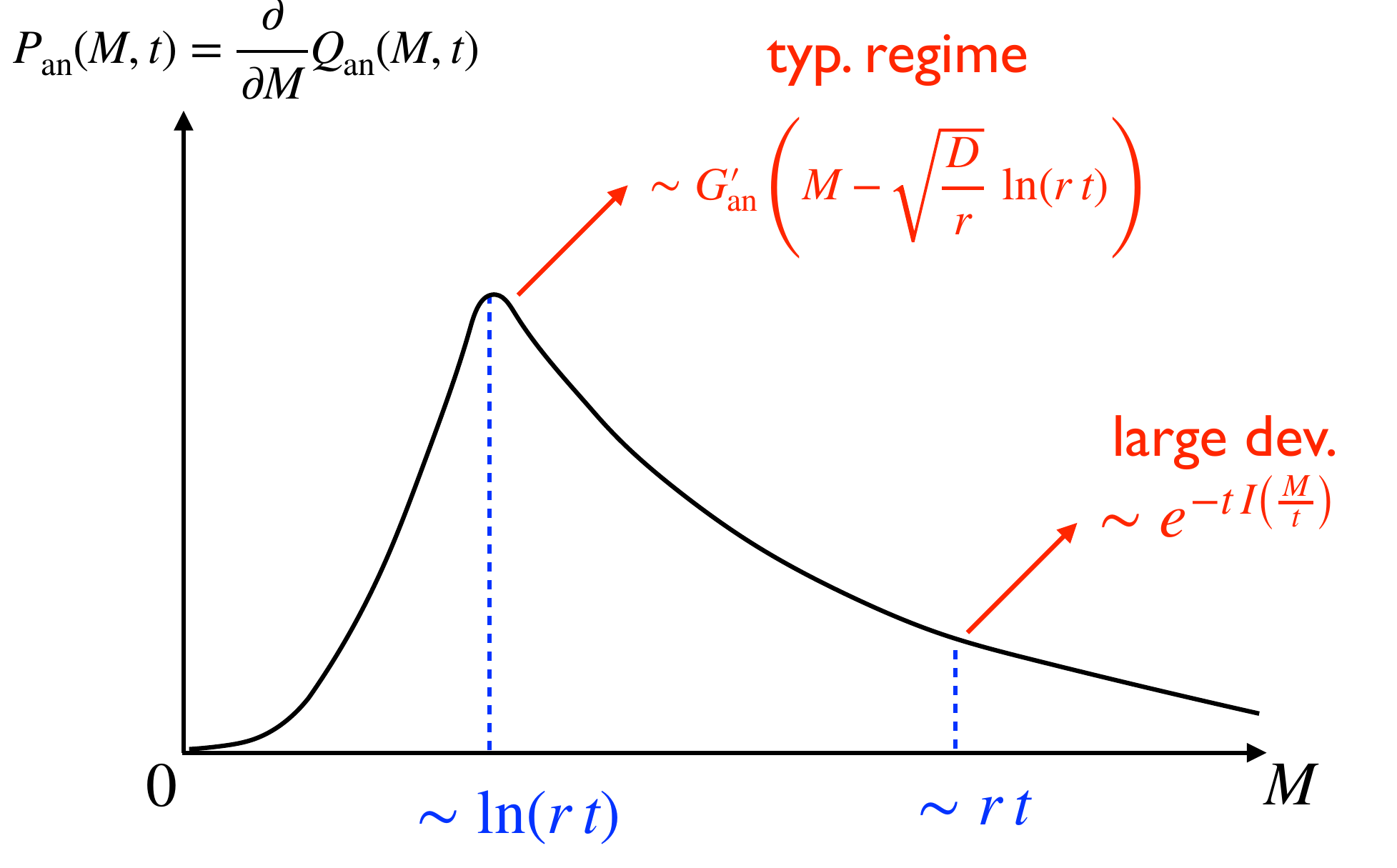}
\caption{A schematic plot of the annealed PDF $P_{\rm an}(M,t)$ vs. $M$ in Eq.~\eqref{PDF_an_summary}. The two vertical dashed blue lines represent the typical and the large-deviation scales.}\label{Fig_N_particle}
\end{figure}

\noindent{\bf Typical regime:} In this case, we anticipate and verify a posteriori that the typical
value of $M_t$ scales as $\sqrt{D/r}\, \ln t$ for large $t$ with $O(1)$ fluctuations around it.
To extract this typical behavior from the Laplace transform in Eq.~(\ref{J_Mt}), we rescale
$s\, t= \tilde{s}$ and re-write it as
\bea
J(M,t) =\sqrt{\frac{D}{r}}\, \int_\Gamma \frac{d\tilde{s}}{2\pi i}\, \frac{e^{\tilde{s}}}{\tilde{s}} \, 
\sqrt{1+ \frac{\tilde{s}}{rt}}\,
\ln \left(1 + \frac{r\, t}{\tilde{s}}\, e^{-\sqrt{\frac{r+\frac{\tilde{s}}{t}}{D} }\, M} \right) \; .
\label{JMt.1}
\eea
Now we take the $t\to \infty$ limit, $M\to \infty$ limit but keeping $M- \sqrt{D/r}\, \ln (r t) =z$ fixed.
One immediately gets in this scaling limit
\bea
J(M,t) \approx F\left(M- \sqrt{\frac{D}{r}}\, \ln (rt)\right)\, , \quad {\rm where}\quad
F(z)=\sqrt{\frac{D}{r}}\, \int_\Gamma \frac{d\tilde{s}}{2\pi i}\, \frac{e^{\tilde{s}}}{\tilde{s}}\, 
\,
\ln \left( 1+ \frac{1}{\tilde{s}}\, e^{-\sqrt{r/D}\, z}\right)\, .
\label{JMt_scaling.1}
\eea 
The scaling function $F(z)$ can be evaluated exactly by 
expanding the logarithm in the integrand in Eq.~(\ref{JMt_scaling.1}) in 
a power series in $1/{\tilde s}$ and then 
performing the Bromwich 
integral term by term. This gives
\bea \label{Fz}
F(z) = - z + \sqrt{\frac{D}{r}}\left(\gamma_E + 
\Gamma(0, e^{- \sqrt{\frac{r}{D}}z})\right) \;,
\eea
where $\Gamma(0,u) = \int_u^\infty \frac{e^{-x}}{x}\,dx$ is the 
incomplete Gamma function. 
In particular, the asymptotic behaviors of $F(z)$ are given by
\bea \label{asympt_F}
F(z) \approx
\begin{cases}
&- z + \sqrt{\frac{D}{r}}\, \gamma_E \quad, \quad\quad\quad\quad z\to - \infty \\
& \\
& \sqrt{\frac{D}{r}}\,  e^{- \sqrt{\frac{r}{D}}z} \quad \quad \hspace*{1.3cm} z \to + \infty \;.
\end{cases}
\eea

Then using Eq.~(\ref{CDF_N_thermo.2}), we finally obtain our main result
\bea \label{CDF_annealed}
Q_{\rm an}(M,t) = e^{- \rho\, F\left(M-\sqrt{\frac{D}{r}}\, \ln (rt)\right)}\, .
\eea
Thus the CDF of $M_t$ has the final scaling form 
\bea \label{CDF_scaling}
Q_{\rm an}(M,t) &\approx& 
G_{\rm an}\left( M - \sqrt{\frac{D}{r}}\, \ln (rt)\right)\, \\
 \quad {\rm where} \quad G_{\rm an}(z) &=& e^{- \rho F(z)} = 
\exp\left[-\rho \sqrt{\frac{D}{r}}\gamma_E + \rho z - 
\rho \sqrt{\frac{D}{r}}\,\Gamma\left(0,e^{-\sqrt{\frac{r}{D}\,z}}\right)   
\right]\; .
\eea
This function $G_{\rm an}(z)$ is a new scaling function, different from the standard 
Gumbel form. 
Note that in the limit of large $M$, i.e., as $z\to \infty$, using the second line in Eq.~(\ref{asympt_F}),
we get
\bea
G_{\rm an}(z) \approx e^{-\rho\, \sqrt{\frac{D}{r}}\, e^{-\sqrt{r/D}\, z}}\, ,
\label{Gumbel.1}
\eea
representing typically the Gumbel-like behavior of the CDF. In the opposite limit when $M\to 0$,
i.e., when $z \to -\infty$, we get from the first line of (\ref{asympt_F})  
\bea \label{survival}
Q_{\rm an}(M\to 0,t) \sim e^{-\rho \sqrt{D/r}\ln (rt)} \sim t^{-\rho\, \sqrt{r/D}}\, , 
\eea
thus recovering the known result (\ref{Qanm0.1}) of Ref.~\cite{EM_11}.
As $M$ increases from $0$, our main new result in Eq.~(\ref{CDF_scaling}) thus 
interpolates smoothly between $M=0$ and
the typical value of $M\sim \sqrt{D/r}\, \ln t$.
Finally, it follows that in this typical regime, the PDF 
\bea
P_{\rm an}(M,t)= \frac{\partial}{\partial M} Q_{\rm an}(M,t)\, ,
\label{an_pdf.1}
\eea
has the scaling behavior
\bea
P_{\rm an}(M,t)\approx G_{\rm an}'\left( M - \sqrt{\frac{D}{r}}\, \ln (rt)\right)\, ,
\label{an_pdf_typ.1}
\eea
where $G'(z)= dG/dz$ is the derivative of the scaling function $G(z)$ in 
Eq.~(\ref{CDF_scaling}).

\vspace*{0.5cm}

\noindent{\bf Large-deviation regime:}
We start from Eq.~(\ref{CDF_N_thermo.2}) where $J(M,t)$ is given explicitly by Eq.~(\ref{J_Mt}).
Note that the result in Eq.~(\ref{J_Mt}) is valid for all $M$. 
For the typical fluctuations, we considered the scaling limit
$M\to \infty$, $t\to \infty$ with $M-\sqrt{D/r}\, \ln t=z$ fixed. 
For the large-deviation regime, when $M$ is very large, one can expand the logarithm 
in (\ref{J_Mt}) and only keep the first term, resulting in
\bea \label{J_Mt.2}
J(M,t) \approx \int_\Gamma \frac{ds}{2\pi i}\, e^{st} \, 
\frac{\sqrt{D(r+s)}}{s^2}e^{-\sqrt{\frac{r+s}{D}}M} \;.
\eea
The integrand has a double pole at $s=0$ and a branch cut at $s=-r$.
In addition it has a saddle point at
$s_1 = -r +  (M/t)^2/{(4D)}$ as in Eq.~(\ref{s1}). 
If $s_1 > 0$, the dominant contribution to the integral comes from the 
saddle point, and evaluating this saddle point one gets the large 
deviation behavior $J(M,t) \approx e^{- r\, t - M^2/(4Dt)}$. The condition
$s_1>0$ translates into $M > M_c(t)$ where
\bea \label{Mc.2}
M_c(t) = \sqrt{4\,D\,r}\,t \;.
\eea
When $M<M_c(t)$, the double pole at $s=0$ dominates and one gets 
$J(M,t) \approx t\, e^{-\sqrt{\frac{r}{D}}M}$. Thus, in the large-deviation regime when $M= O(t)$, 
we can combine the behaviors for $M<M_c(t)$ and $M>M_c(t)$ into a single large-deviation form
\bea \label{J_sum}
J(M,t) \approx e^{-t \, I\left(\frac{M}{t}\right)} \quad, \quad I(z) = 
\begin{cases}
&\sqrt{\frac{r}{D}}\,z\, ,  \quad\, z < z_c=\sqrt{4\,D\,r} \\
&\\
& r + \frac{z^2}{4D}\, , \quad z > z_c= \sqrt{4\,D\,r} \;.
\end{cases}
\eea
The rate function $I(z)$ is identical to the single
particle case in Eq.~(\ref{rate_function}) and
thus exhibits a second-order dynamical phase transition at $z = \sqrt{4\,D\,r}$. Hence, in this large-deviation
regime, using Eqs. (\ref{CDF_N_thermo.2}) and (\ref{J_sum}) we get the annealed CDF
\bea \label{Q_sum}
Q_{\rm an}(M,t) \approx \exp{\left[-\rho \, e^{-t\,  I\left(\frac{M}{t}\right)}\right]}
\approx 1- \rho\, e^{-t\,  I\left(\frac{M}{t}\right)} \; .
\eea

Another alternative way to analyze the large-deviation regime is by probing directly
the annealed PDF in Eq.~(\ref{an_pdf.1}).
Taking derivative of Eq.~(\ref{CDF_N_thermo.2}) with respect to $M$ gives the exact expression
\bea
P_{\rm an}(M,t)= \rho\, \left[1- Q_r(M,t)\right]\, e^{-\rho\, J(M,t)}\, .
\label{an_pdf.2}
\eea
We now substitute in Eq.~(\ref{an_pdf.2}) the large-deviation behavior of $Q_r(M,t)$ in Eq.~(\ref{2_3_Pr}) and
that of $J(M,t)$ in Eq.~(\ref{J_sum}). This leads for large $t$, large $M$ but with $M/t$ fixed, the
following large-deviation behavior of $P_{\rm an}(M,t)$ (up to pre-exponential factors)
\bea \label{P_sum}
P_{\rm an}(M,t) \sim e^{-t\, I\left(\frac{M}{t}\right)}\, ,
\eea
with the rate function $I(z)$ given in Eq.~(\ref{rate_function}).

Thus, to summarize, we find that in the annealed case, there are essentially two scales of $M$, namely
$M\sim \sqrt{D/r}\,\ln (rt)$ characterizing typical fluctuations and $M\sim t$ characterizing
atypically large fluctuations. The behavior of the PDF of the maximum, for large $t$, can
be summarized as 
\bea
P_{\rm an}(M,t)\approx \begin{cases}
& G_{\rm an}'\left( M - \sqrt{\frac{D}{r} }\, \ln (rt)\right)\, , 
\quad {\rm when} \quad M\sim \sqrt{\frac{D}{r} }\, \ln (rt)\, , \\
& \\
& e^{-t\, I\left(\frac{M}{t}\right)}\, , \quad\quad\quad\quad\quad\quad\quad\,\, {\rm when} \quad M\sim t\, ,
\end{cases}
\label{PDF_an_summary}
\eea
with $G(z)$ and $I(z)$ given respectively in Eqs. (\ref{CDF_scaling}) and (\ref{rate_function})
A schematic plot of $P_{\rm an}(M,t)$, in the typical and in the large-deviation regimes, is shown 
in Fig.~\ref{Fig_N_particle}.

\subsection{Quenched case}

In the quenched case, the survival probability or equivalently, the CDF $Q_{\rm qu}(M,t)$ is given by
Eq.~(\ref{qu_avg.1}), where
the average $\overline{ \cdots}$ is again over the initial positions $\{x_i \}$'s, 
each drawn independently and uniformly from $x_i \in
[-L,0]$. Performing this average (by changing $x_i\to -x_i$) and taking the thermodynamic limit gives
\begin{equation}
Q_{\rm qu}(M,t)= \exp\left[ \overline{\ln Q_r(M,t|\{x_i\})}\right]= \exp\left[ \frac{1}{L}\, \sum_{i=1}^N 
\int_{-L}^0 \ln \left[ Q_r(M-x_i, t)\right]\, dx_i\right]= \exp\left[\rho\, \int_0^{\infty} 
\ln \left[Q_r(M+x,t)\right]\, dx\right]\, .
\label{qu_cdf.1}
\end{equation}
Shifting $x$ by $M+x$ in the integrand gives
\bea
Q_{\rm qu}(M,t)= \exp\left[ \rho\, \int_M^{\infty} \ln\left[Q_r(m,t)\right]\, dm\right]\, ,
\label{qu_cdf.2}
\eea
where we recall that $Q_r(m,t)$ is the single particle survival probability with a target fixed at $m$, 
and is given by Eq.~(\ref{Brom_Qr}), i.e.,
\bea 
\label{Brom_Qr.1}
Q_r(m,t) = \int_{\Gamma} \frac{ds}{2\pi i} \,e^{s\,t}\,  
\frac{1 - e^{-\sqrt{\frac{r+s}{D}}\, m}}{s + r\, e^{-\sqrt{\frac{r+s}{D}\,}\, m}}
\eea
Once again, the special case $M=0$, i.e., the quenched survival probability $Q_{\rm qu}(M=0,t)$ for 
a target fixed at the origin was studied in Ref.~\cite{EM_11} and it was found to decay exponentially
for large $t$ as
\bea
Q_{\rm qu}(M=0,t) \approx \exp\left[- \rho\, \sqrt{r\, D}\, \theta_1\, t\right]\, , 
\quad {\rm with}\quad \theta_1= 4 (1-\ln 2)\, .
\label{qu_m0.1}
\eea
Note that in Ref.~\cite{EM_11}, the exponent was $2\,\theta_1$ since the effective density there
was $2\rho$.
Our goal here is to study $Q_{\rm qu}(M,t)$ for all $M\ge 0$, going beyond the special case $M=0$.

\begin{figure}
\includegraphics[width = 0.7\linewidth]{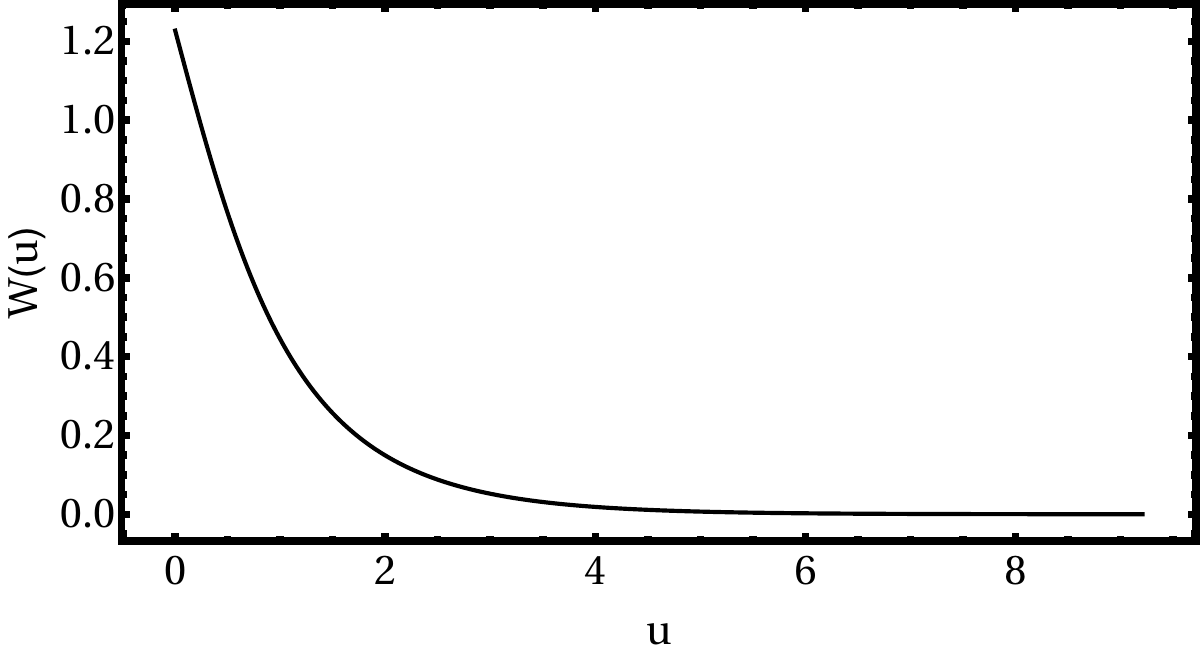}
\caption{The function $W(u)$ vs. $u$ obtained by eliminating $y^*(u)$ between Eqs. (\ref{Wu.1})
and (\ref{ystaru.1}) using Mathematica. The function $W(u)\to 4 (1-\ln 2)=1.22741\cdots$ as $u\to 0$
and decays exponentially as $W(u)\approx e^{-u}$ as $u\to \infty$ as described in Eq.~(\ref{Wu_asympt}).}
\label{Fig_Wu}
\end{figure}

To make progress for general $M\ge 0$, we first note that for large $t$, the dominant behavior
of the integral in Eq.~(\ref{Brom_Qr.1}) is governed by the pole of the integrand and one has, 
to leading order for large $t$ (ignoring pre-exponential factors),
\bea
Q_r(m,t) \approx e^{s^*(m)\, t}\, ,
\label{Qr_pole.1}
\eea
where the pole $s^*(m)$ satisfies 
\bea
s^*(m) + r\, e^{-\sqrt{\frac{\left(r+s^*(m)\right)}{D}}\, m}=0 \, .
\label{pole.1}
\eea
Substituting (\ref{Qr_pole.1}) in Eq.~(\ref{qu_cdf.2}) gives, up to pre-exponential factors,
\bea
Q_{\rm qu}(M,t)\approx \exp\left[ \rho\, t\, \int_M^{\infty} s^*(m)\, dm\right]\, .
\label{qu_cdf.3}
\eea
The pole $s^*(m)$ in Eq.~(\ref{pole.1}) is necessarily negative. It is then natural
to substitute $s^*(m)= - r\, y$ in Eq.~(\ref{qu_cdf.3}) such that $y\ge 0$. Under this substitution,
it follows from Eq.~(\ref{pole.1}) that the variable $y$ satisfies
\bea
y= e^{-\sqrt{\frac{r}{D}\, (1-y)}\, m}\, , \quad {\rm implying}\quad m= 
-\sqrt{\frac{D}{r}}\, \frac{\ln y}{\sqrt{1-y}}\, .
\label{pole.2}
\eea
We now make the change of variable $m\to y$ in Eq.~(\ref{qu_cdf.3}) and obtain, using (\ref{pole.2})
\bea
Q_{\rm qu}(M,t)\approx \exp\left[ -\rho\, \sqrt{r\, D}\, t\, W\left(\sqrt{\frac{r}{D}}\, M\right)\right]\, .
\label{qu_cdf.4}
\eea
Here the function $W(u)$ is given by
\bea
W(u)= \int_0^{y^*(u)} \left[\frac{1}{\sqrt{1-y}}+ \frac{y\, \ln y}{2\, (1-y)^{3/2}}\right]\, dy\, ,
\label{Wu_def}
\eea
where $0\le y^*(u)\le 1$ is determined in terms of $u$ by solving the equation
\bea
y^*(u)= e^{-\sqrt{1-y^*(u)}\,u} \, . 
\label{ystaru.1}
\eea
It turns out that the integral in Eq.~(\ref{Wu_def}) can be performed explicitly, giving
\bea
W(u)= 4\, \left(1- \sqrt{1-y^*}\right) + 4\, \left[{\rm arctanh} \left(\sqrt{1-y^*}\right)-\ln 2 \right]+ 
\frac{ (2-y^*)\, \ln (y^*)}{\sqrt{1-y^*} }\, ,
\label{Wu.1}
\eea
with $y^*\equiv y^*(u)$ determined from Eq.~(\ref{ystaru.1}) as a function of $u$. From Eq.~(\ref{ystaru.1}), it
is easy to derive the asymptotic behavior of $y^*(u)$. One gets
\bea
y^*(u)\approx \begin{cases}
& 1-u^2\, ,  \quad {\rm as} \quad  u\to 0  \\
&\\
& e^{-u} \, , \quad\quad {\rm as} \quad u\to \infty\, .
\end{cases}
\label{yu_asympt}
\eea 
Consequently, the asymptotic behaviors of $W(u)$ in Eq.~(\ref{Wu.1}) are given by
\bea
W(u) \approx \begin{cases}
& 4\, (1-\ln 2) - u \, ,  \quad {\rm as} \quad  u\to 0  \\
&\\
& e^{-u} \, , \quad\quad\quad\quad\quad\quad {\rm as} \quad u\to \infty\, .
\end{cases}
\label{Wu_asympt}
\eea
A plot of the function $W(u)$, obtained using Mathematica, is provided in Fig.~\ref{Fig_Wu}.
Substituting the $W(u\to 0)$ behavior above in Eq.~(\ref{qu_cdf.4}), we reproduce the $M\to 0$ limit
of $Q_{\rm qu}(M=0,t)$ in Eq.~(\ref{qu_m0.1}). On the other hand, 
substituting the large $u$ behavior of $W(u)\approx e^{-u}$ in Eq.~(\ref{qu_cdf.4}), we see that
a natural scaling limit emerges: $M\to \infty$, $t\to \infty$ but with $M-\sqrt{D/r}\, \ln (rt)=z$ fixed.
This is indeed the typical behavior of $Q_{\rm qu}(M,t)$.

\vspace*{0.5cm}

\noindent{ {\bf Typical regime:}} In this typical scaling regime, where $M$ and $t$ are both large
with $M-\sqrt{D/r}\, \ln (rt)=z$ fixed, we then get from Eq.~(\ref{qu_cdf.4})
\bea
Q_{\rm qu}(M,t) \approx G_{\rm qu}\left(M- \sqrt{\frac{D}{r}}\, \ln (rt)\right)\, ,
\quad {\rm where}\quad G_{\rm qu}(z)= 
\exp\left[- \rho\, \sqrt{\frac{D}{r}}\, e^{-\sqrt{r/D}\, z}\right]\, .
\label{Qu_typ.1}
\eea
One should compare this typical scaling behavior in the quenched case with that of the annealed
case in Eq.~(\ref{CDF_scaling}). While in the annealed case, the scaling function
$G_{\rm an}(z)$ is a new function not of the Gumbel form, in the quenched case
the scaling function $G_{\rm qu}(z)$ in Eq.~(\ref{Qu_typ.1})  has a Gumbel form. 
Another difference between the two cases is worth remarking. In the annealed case,
the very small $M$ behavior ($M\to 0$) is included in the scaling form in Eq.~(\ref{CDF_scaling})
corresponding to the $z\to -\infty$ limit. In contrast, in the quenched case, the $M\to 0$
limit in Eq.~(\ref{qu_m0.1}) is not part of the scaling function $G_{\rm qu}(z)$
in Eq.~(\ref{Qu_typ.1}). To see this, we note that substituting $M=0$, i.e.,
$z=-\sqrt{\frac{D}{r}}\, \ln (rt)$ in Eq.~(\ref{Qu_typ.1}) gives, $Q_{\rm qu}(0,t)\approx
\exp[-\rho\, \sqrt{D\, r}\, t] $ which is different from the true $M=0$ behavior
in Eq.~(\ref{qu_m0.1}). In the quenched case, the behavior of $Q_{\rm qu}(M,t)$, for both small
and large $M\sim O(\ln(rt))$, is rather captured by the result in Eq.~(\ref{qu_cdf.4})
with the function $W(u)$ given explicitly in Eqs. (\ref{Wu.1}) and (\ref{ystaru.1}).

\vspace*{0.5cm}

\noindent{\bf Large-deviation regime:} In the discussion about typical fluctuations above we probed the 
CDF $Q_{\rm qu}(M,t)$ as a function of $M$, for fixed large $t$, at the scale 
$M\sim \sqrt{D/r}\, \ln t$. Let us now investigate the same CDF $Q_{\rm qu}(M,t)$
at the scale of large deviations when $M\sim O(t)$. As in the annealed case, to investigate
this large-deviation regime, it is convenient to consider the PDF of $M$, namely,
\bea
P_{\rm qu}(M,t)= \frac{\partial}{\partial M} Q_{\rm qu}(M,t)\, .
\label{qu_pdf.1}
\eea
Taking derivative of Eq.~(\ref{qu_cdf.2}) with respect to $M$ yields the exact PDF
\bea
P_{\rm qu}(M,t)=-\rho\, \ln\left[Q_r(M,t)\right]\, \exp\left[ \rho\, \int_M^{\infty} 
\ln\left[Q_r(m,t)\right]\, dm\right]\, ,
\label{qu_pdf.2}
\eea
where $Q_r(m,t)$ is the single particle CDF given in Eq.~(\ref{Brom_Qr.1}). In the large-deviation regime
when $M\sim O(t)$, we then use the large-deviation result for $Q_r(M,t)$ in Eq.~(\ref{2_3_Pr}).
Substituting this result in Eq.~(\ref{qu_pdf.2}) and taking the limit $t\to \infty$, it is
straightforward to see that the exponential factor in Eq.~(\ref{qu_pdf.2}) contributes $1$
to leading order and expanding the logarithm in the front one gets, to leading order for large $t$ (ignoring
pre-exponential factors as usual)
\bea
P_{\rm qu}(M,t)\approx e^{-t\,  I\left( \frac{M}{t}\right)}\, ,
\label{qu_pdf_ldv.1}
\eea
where the rate function $I(z)$ is given exactly in Eq.~(\ref{rate_function}). 
The second derivative of the rate function $I(z)$ is discontinuous at $z=z_c=\sqrt{4Dr}$,
signaling a second-order phase transition. The result for the quenched case in Eq.~(\ref{qu_pdf_ldv.1})
is identical (up to pre-exponential factors) to that in the annealed case in Eq.~(\ref{P_sum}). 
Thus, on the scale $M\sim O(t)$,
both the annealed and the quenched PDF's of $M$, respectively in Eqs. (\ref{P_sum})
and (\ref{qu_pdf_ldv.1}), are described by an identical large-deviation form that coincides with
the single particle case in Eq.~(\ref{2_3}). 

\begin{figure}[t]
\includegraphics[angle=-0,width = 0.6\linewidth]{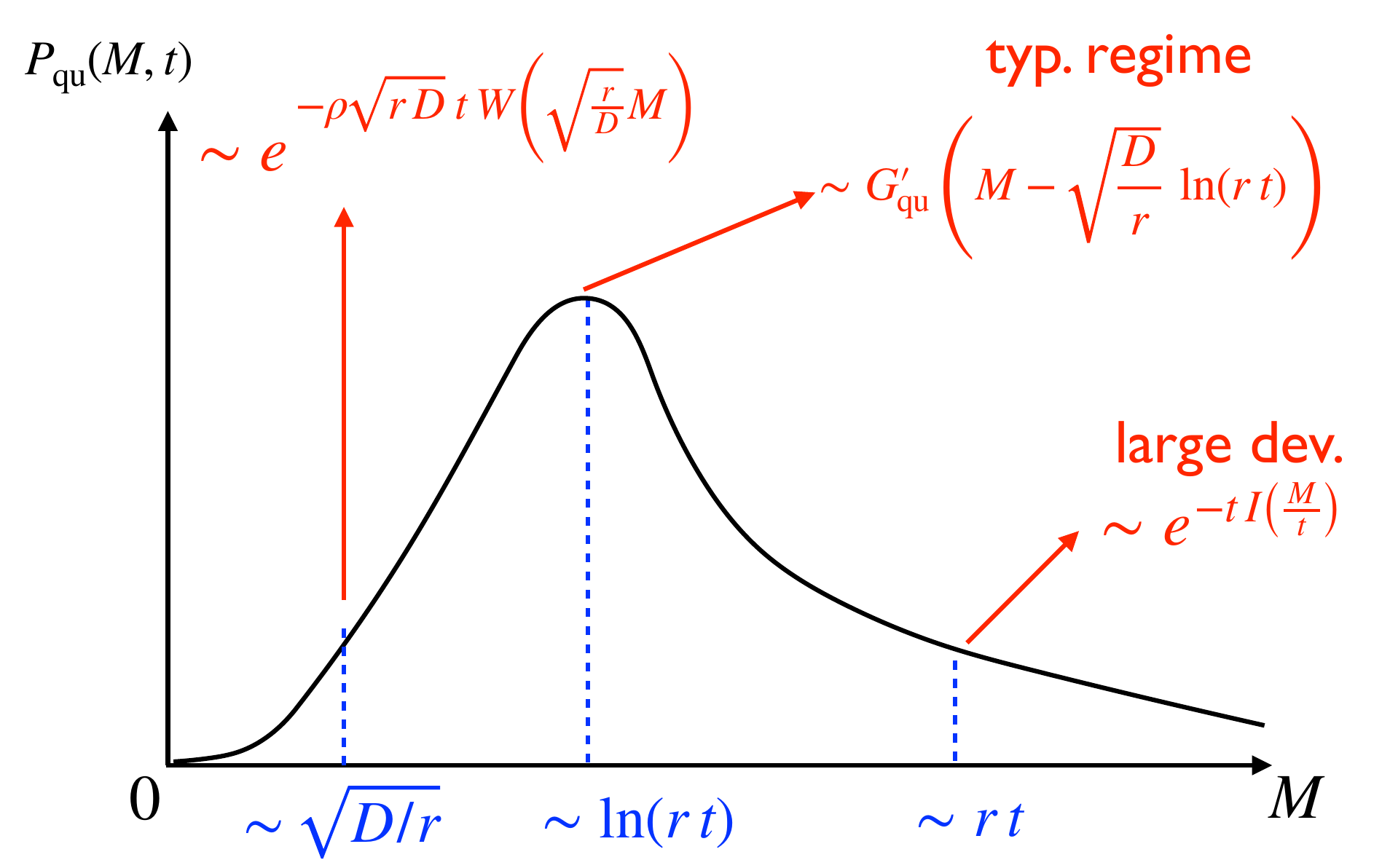}
\caption{\Blue{A schematic plot of the quenched PDF $P_{\rm qu}(M,t)$ vs. $M$ in Eq.~\eqref{PDF_qu_summary}. The three vertical dashed blue lines represent the three scales mentioned in Eq. (\ref{PDF_qu_summary}).}}\label{Fig_N_particle_qu}
\end{figure}

Thus, in the quenched case, the behavior of $Q_{\rm qu}(M,t)$ or equivalently the PDF $P_{\rm qu}(M,t)= \partial_M
Q_{\rm qu}(M,t)$ can be summarized as follows (ignoring pre-exponential factors as usual) 
\bea
P_{\rm qu}(M,t) \approx \begin{cases}
& \exp\left[ -\rho\, \sqrt{r\, D}\, t\, W\left(\sqrt{\frac{r}{D}}\, M\right)\right]\, , 
\quad\, {\rm when}\quad M\sim \sqrt{\frac{D}{r}}\, ,  \\
& \\
& G_{\rm qu}'\left(M- \sqrt{\frac{D}{r}}\, \ln (rt)\right)\, , \quad\quad\quad\,\,\, {\rm when} 
\quad M\sim \sqrt{\frac{D}{r}}\, \ln (rt)\, , \\
& \\
&  e^{-t\,  I\left( \frac{M}{t}\right)}\, , \quad\quad\quad\quad\quad\quad\quad\quad\quad\quad {\rm when}\quad M\sim t\, ,
\end{cases}
\label{PDF_qu_summary}
\eea
where the functions $W(u)$, $G_{\rm qu}(z)$ and $I(z)$ are given respectively in Eqs. (\ref{Wu.1}),
(\ref{Qu_typ.1}) and (\ref{rate_function}). A schematic plot of the quenched PDF $P_{\rm qu}(M,t)$ is given in Fig. \ref{Fig_N_particle_qu}. 
Thus, compared to the annealed case in Eq.~(\ref{PDF_an_summary}),
we see that in the quenched case there is an additional scale when $M\sim O(1)$ for large $t$, where the PDF
$P_{\rm qu}(M,t)$ behaves differently from the typical regime. This is natural because in the quenched case
a scale of $O(1)$ naturally emerges from the initial condition where the particles are equispaced.

\section{Numerical Simulations}

To study the problem numerically \cite{practical_guide2015}, 
we start with the {\bf single particle} case.
To obtain a numerical approximation of the cumulative 
 distribution $Q_r(M,t)$ at a fixed given time $t$, 
we considered an array $\overline{Q}(M,t')$
which is discretised in time with time resolution $\Delta t$, i.e.,
for $t'=0,\Delta t, 2\Delta t, \ldots, t$. We
applied a discretised version of Eq.~(\ref{renewal}) which reads

\begin{equation}
\overline{Q}(M,t') = e^{-rt'}Q_0(M,t')+ (1-e^{-r\Delta t})
\sum_{\tau=\Delta t, 2\Delta t,\ldots }^{t'}
 e^{-r (\tau-\Delta t)}\overline{Q}(M,t'-\tau) Q_0(M,\tau)\,
\label{eq:Q:numeric}
\end{equation}
where the survival probability $Q_0(M,t)$ of an 
ordinary Brownian motion without resetting is given by Eq.~(\ref{survival_BM})
and
$e^{-r}$ is the probability of experiencing no reset in the 
unit time span. Thus, each term in the sum represents the case
that a particle experiences the last reset at time $t'-\tau$, i.e., 
$\tau/\Delta t$ ``steps'' back in time,
for all possible values of $\Delta t \le \tau \le t'$. 
This happens with
the  probability which is a product of the probability 
$(1-e^{-r\Delta t})$ for the  last reset times
the probability $e^{-r(\tau-\Delta t)}$ for 
experiencing no reset since the last reset, during time $\tau-\Delta t$.
 For the calculation of
$\overline{Q}(M,t')$ one includes the probabilities $\overline{Q}(M,t'-\tau)$ 
and $Q_0(M,\tau)$ that a particle
does not reach position $M$ in the both 
time intervals before and after the last reset, respectively.

To achieve a high
numerical accuracy in
evaluating very small probabilities like $100^{-100}$ and
estimating the sums and differences of probabilities,
we used the \verb!mpfr! library \cite{mpfr-lib,fousse2007} with a 
precision of 1000
binary digits for storing data in arrays and performing arithmetic operations.
 
To estimate numerically the probability density,
we approximate Eq.~(\ref{max.3})
by a finite difference
\begin{equation}
P_r(M,t) \approx \frac{\overline{Q}_r(M+h,t)-\overline{Q}_r(M-h,t)}{2h}\,.
\label{eq:difference}
\end{equation}
Thus, to investigate $P_r(M,t)$  for a sequence $M=k\Delta M$ 
($k=1,2,3,\ldots)$ of values of $M$,
we actually considered the values $\Delta M-h,\Delta M+h,2\Delta M-h,
2\Delta M+h,3\Delta M-h, \ldots$.
We tested different spacing values $h$ and found no notable changes
when going below $h=0.01$.

\begin{figure}
\includegraphics[width = 0.5\linewidth]{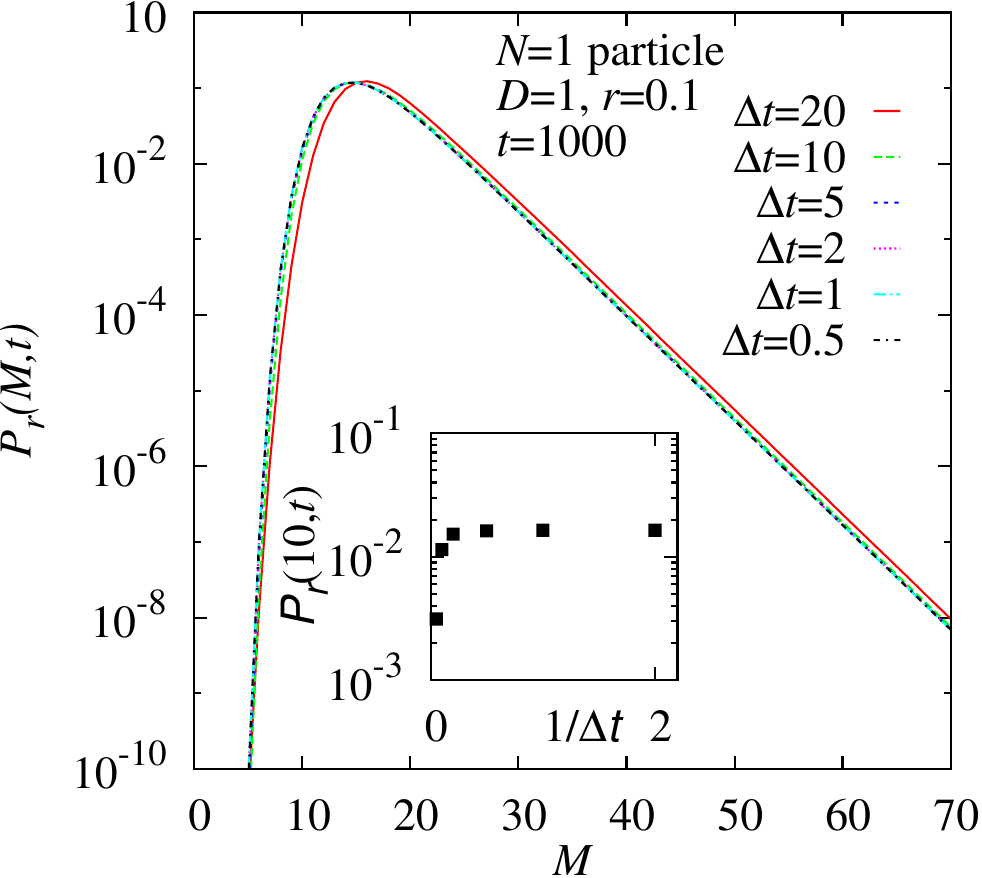}
\caption{One-particle distribution 
$P_{r}(M,t)$ for $t=1000$, $D=1$, $r=0.1$
and various values $\Delta t$ of the discretisation in time. The inset
shows how the particular value $P_r(10,t)$ depends on $1/\Delta t$.
\label{fig:Pr_M_test}}
\end{figure}

Note that for $\Delta t\to 0$
the factor in front of the sum in Eq.~(\ref{eq:Q:numeric})
becomes $r\Delta t$ and 
$e^{-r(\tau-\Delta t)} \to e^{-r\tau}$, therefore
one obtains  Eq.~(\ref{renewal}). One can expect an influence of the
discretisation but it will be small, because quantities like the
survival probability $Q_0(M,\tau)$ and the no-reset
probability $e^{-rt'}$ are always exact independently of the choice of
$\Delta t$. We
have tested this explicitly by calculating the single-particle distribution
$P_r(M,t)$
for various values of $\Delta t$, here for the parameter values 
$t=1000$, $D=1$ and $r=0.1$. The result is shown in Fig.~\ref{fig:Pr_M_test}.
For values below $\Delta t=2$ we observed no significant change in the
results in Fig.~\ref{fig:Pr_M_test}, so we kept $\Delta t=1$ 
throughout our study.

\begin{figure}
\includegraphics[width = 0.5\linewidth]{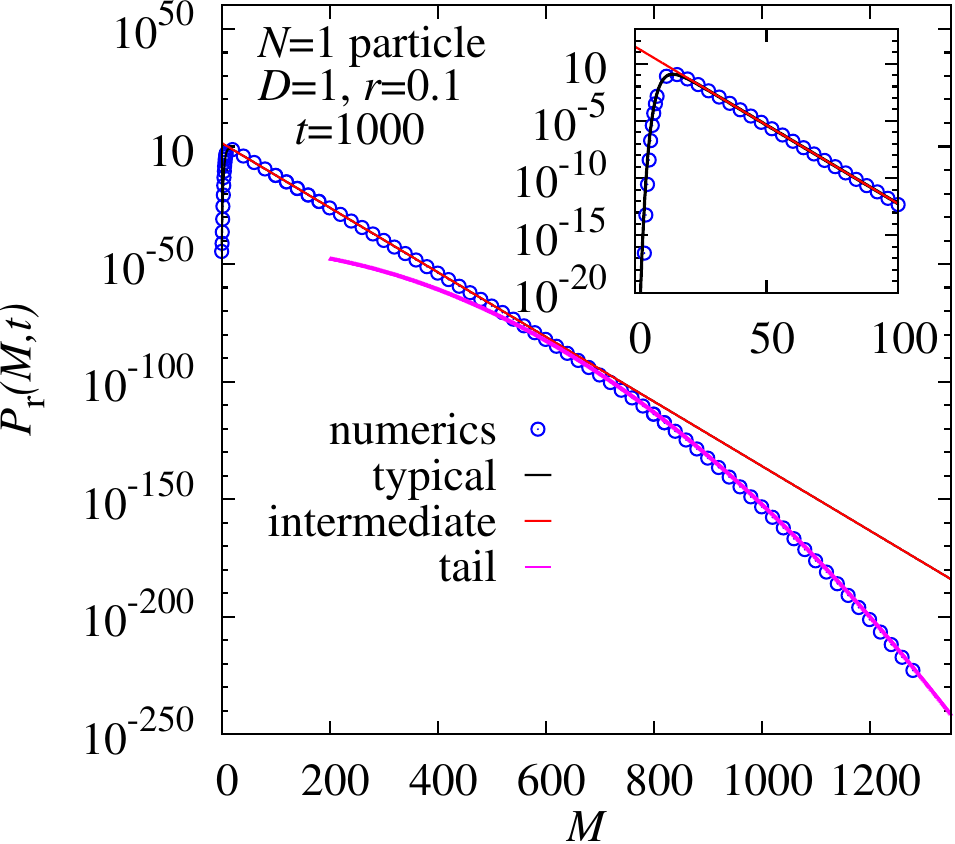}
\caption{Numerical one-particle distribution 
$P_{r}(M,t)$ for $t=1000$, $D=1$, $r=0.1$ (symbols)
compared to the analytical result in Eq.~(\ref{summary_PDF}) 
(lines). The inset highlights the
range of small values of $M$. 
\label{fig:Pr_M_single}}
\end{figure}

\begin{figure}
\includegraphics[width = 0.5\linewidth]{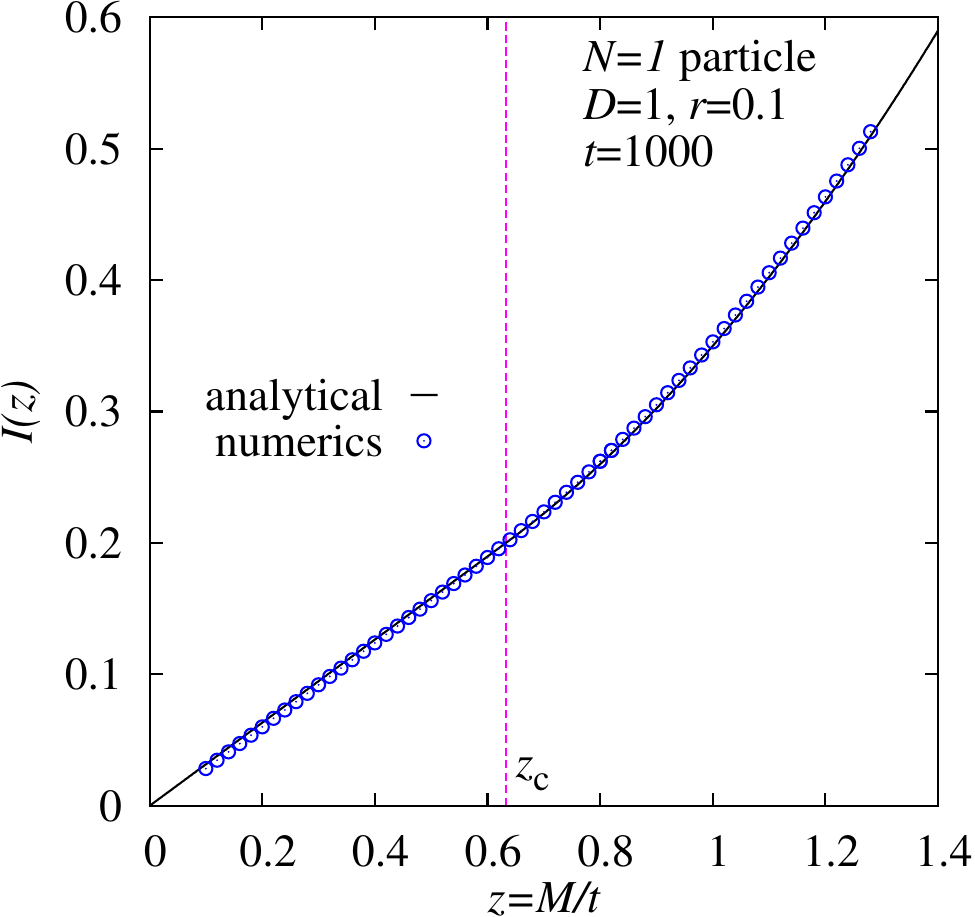}
\caption{Numerical result for the one-particle 
rate function $I(z)$ for $t=1000$, $D=1$, $r=0.1$ 
(symbols) compared to the analytical result of Eq.~(\ref{rate_function}) 
(solid line). The red vertical line provides the location of the critical value $z_c = \sqrt{4D\,r} = 0.632456\ldots$.
\label{fig:Iz}}
\end{figure}

Having tested the dependence on $\Delta t$ of $P_m(M,t)$ in
Fig.~\ref{fig:Pr_M_test}, we then fix $\Delta t=1$ and 
present the result for the single-particle distribution $P_r(M,t)$
vs. $M$ in
Fig.~\ref{fig:Pr_M_single} for $t=1000$ and compared it to the 
analytical result
in Eq.~(\ref{summary_PDF}). Generally, a very good agreement can be
observed, for the left tail, the typical region, the intermediate region
and for the large-deviation Gaussian tail.
The predicted large-deviation behavior
in Eq.~(\ref{rate_function}) is confirmed by plotting the numerically
obtained rate function
$I(z)=-\log(P_r(M,t))/t$
as a function of $z=M/t$ in Fig.~\ref{fig:Iz}.

For the multi-particle {\bf annealed case}
we approximate the average in Eq.~(\ref{an_avg.2}) of
the single-particle cumulative function $Q_r(M,t)$ over
all possible initial position in the interval $[-L,0]$ 
by a finite sum over $N$ evenly-spaced particle positions
$x=-(N-1)\delta, -(N-2)\delta,\ldots,\delta,0$. The particles
exhibit a density $\rho=N/L$, i.e., the spacing is $\delta=L/N$. 
In the present work we use $\delta=1$.
Since starting
at position $-i\delta$ and reaching a maximum $M$ is equivalent to
start at position 0 with reaching a maximum $M+i\delta$, one obtains
\begin{equation}
Q_{\rm avg}(M,t) \approx \frac 1 N \sum_{i=0}^{N-1} Q_r(M+i\delta,t)\,.
\end{equation}
Note that the spacing of the considered values of $M$ must be such that it
is compatible with the needed values $M+i\delta$. 
For any target value of $M$, we also perform the full calculation for $M-h$ and 
$M+h$.

The cumulative distribution in Eq.~(\ref{CFD_N.2}) for the annealed case
is simply $[Q_{\rm avg}(M,t)]^N$ from Eq.~(\ref{an_avg.1}), therefore 
$P_{\rm an}(M,t)=\frac{\partial}{\partial M} [Q_{\rm avg}(M,t)]^N$
$= N \frac{\partial Q_{\rm avg}(M,t)}{\partial M} [Q_{\rm avg}(M,t)]^{N-1}$.  
The partial derivative is again obtained 
by taking the numerical difference in Eq.~(\ref{eq:difference}). The result
for the numerical density is then given by
\begin{equation}
P_{\rm an}(M,t)\approx N\frac{Q_{\rm avg}(M+h,t)-Q_{\rm avg}(M-h,t)}{2h}
[Q_{\rm avg}(M,t)]^{N-1}
\end{equation}

\begin{figure}
\includegraphics[width = 0.5\linewidth]{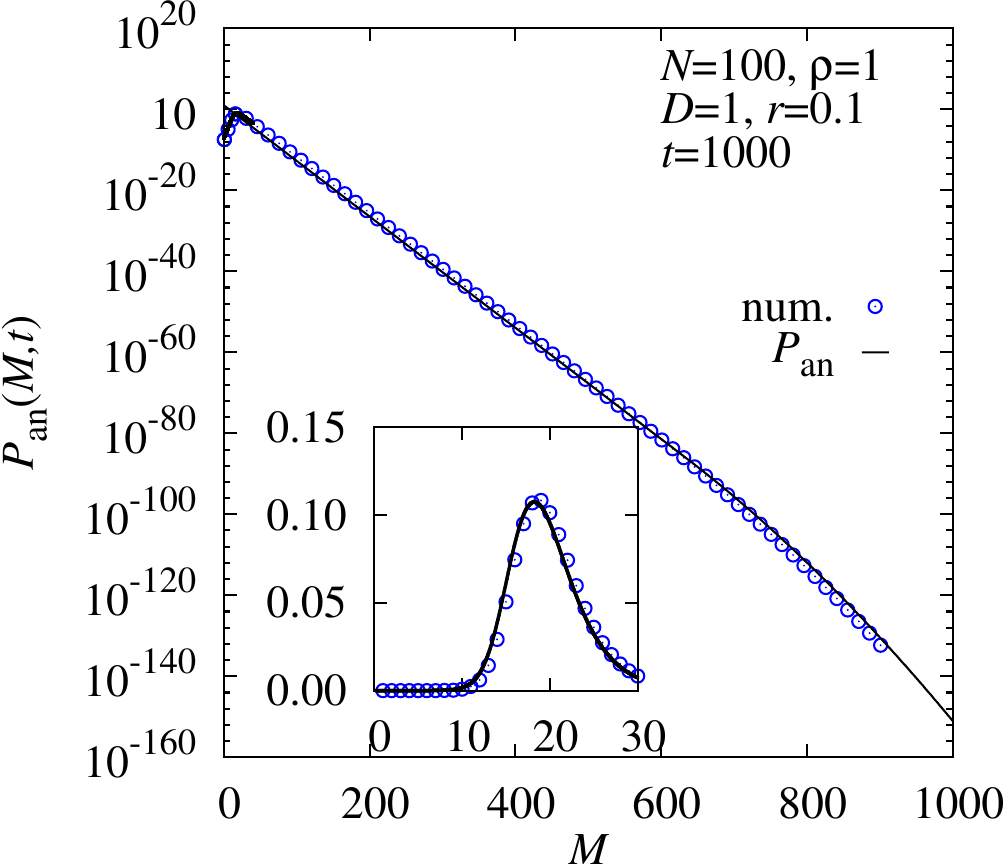}
\caption{Numerical result for the annealed distribution 
$P_{\rm an}(M,t)$ for $t=1000$, $D=1$, $r=0.1$
and $N=100$ particles (symbols). The lines show the analytical result in
Eq.~(\ref{PDF_an_summary}). The inset highlights the typical region in the first line of Eq.~\eqref{PDF_an_summary}.
\label{fig:Pr_M_annealed}}
\end{figure}

The resulting distribution for $N=100$ particles in interval $[-(N-1),0]$
with spacing $\delta=1/\rho=1$
is shown in 
Fig.~\ref{fig:Pr_M_annealed}. We tested that increasing the number $N$
of particles does not change the result, the additional particles are just
too far away. The numerical results confirm very well the analytical
results in Eq.~(\ref{PDF_an_summary}), 
for the typical region, as well as for the tail
which is governed again by the rate function $I(z)$.

For the multi-particle {\bf quenched case}, we consider just one configuration
of particles at positions $x=-(N-1)\delta, -(N-2)\delta,\ldots,\delta,0$. Therefore,
the cumulative distribution is given by
\begin{equation}
Q_{\rm qu}(M,t) = \prod_{i=0}^{N-1} Q_{r}(M+i\delta,t)\,.
\label{num_qu.1}
\end{equation}
To obtain the density, we take the derivative with respect to $M$ and
approximate the derivative of the single-particle cumulative distribution
again by a finite difference arriving at
\begin{equation}
P_{\rm qu}(M,t) = \sum_{i=0}^{N-1}   
\frac{Q_{r}(M+i\delta+h,t)-Q_{r}(M+i\delta-h,t)}{2h}
\prod_{j=0,j\neq i}^{N-1} Q_{r}(M+j\delta,t)\,.
\label{num_qu.2}
\end{equation}

\begin{figure}
\includegraphics[width = 0.5\linewidth]{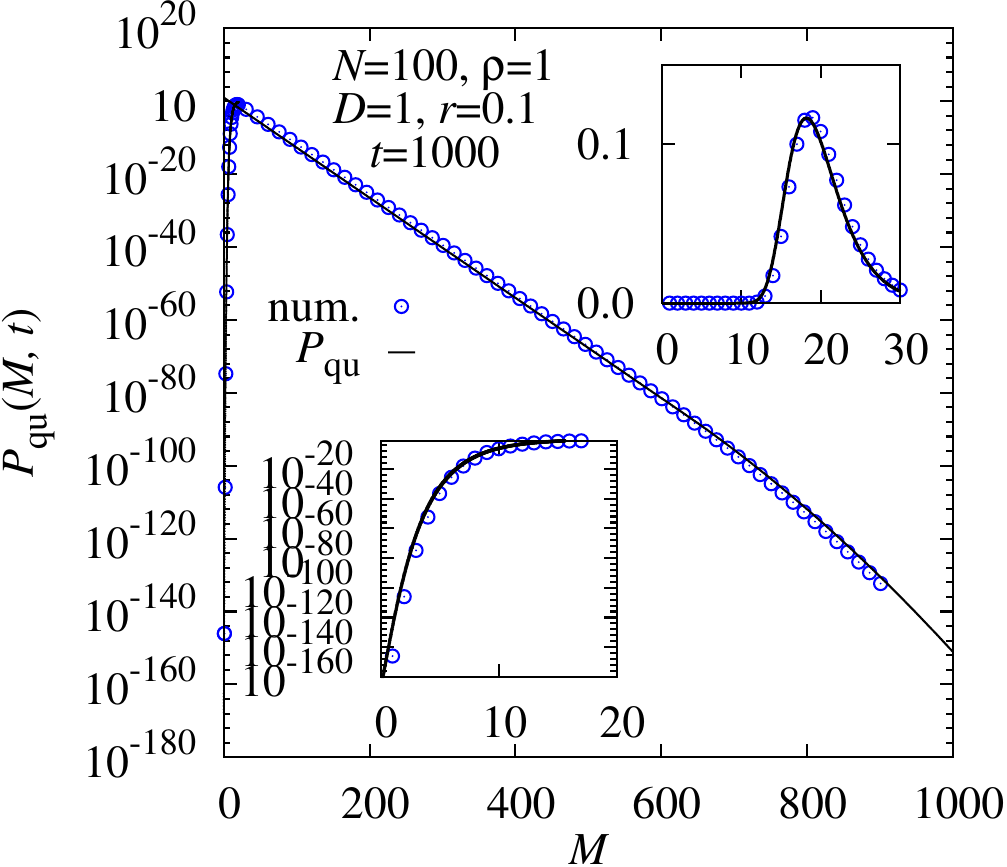}
\caption{Numerical result for the quenched distribution 
$P_{\rm qu}(M,t)$ for $t=1000$, $D=1$, $r=0.1$
and $N=100$ particles (symbols). The lines show the analytical result in
Eq.~(\ref{PDF_qu_summary}). The inset in the upper right corner highlights the
typical regime (as given by the second line of Eq.~\eqref{PDF_qu_summary}). The
inset in the lower left corner corresponds to the left tail, when $M \sim \sqrt{D/r}$, as given in the
first line of Eq.~\eqref{PDF_qu_summary}.
\label{fig:Pr_M_quenched}}
\end{figure}

The resulting distribution for the quenched case
for $N=100$ particles in interval $[-(N-1),0]$
with density $\rho=1$
is shown in 
Fig.~\ref{fig:Pr_M_quenched}. Also here, 
the numerical results confirm very well the analytical
results in Eq.~(\ref{PDF_qu_summary}), for the typical region, 
for both tails and for the intermediate region.

\section{Conclusion}

In this paper, we first revisited the exact computation of the 
probability distribution of the maximum $M_t$ of a single resetting 
Brownian motion (RBM) of duration $t$, starting and resetting at the 
origin with a constant rate $r$. In this case, it was known that the 
random variable $M_t$ fluctuates around $\sqrt{D/r}\,\ln (r\,t)$ for 
large $t$ and the typical scale of fluctuations is of order $O(1)$. The 
appropriately centered and scaled distribution of $M_t$ approaches at 
large times to a standard Gumbel distribution. In this paper, we have 
analyzed the large-deviation regime of $M_t$ when $M_t \sim O(t)$ and 
shown that its PDF satisfies a large-deviation form $P_r(M,t) \sim e^{-t 
I(M/t)}$ where the rate function $I(z)$ has a second-order discontinuity 
at the critical value $z_c = \sqrt{4D\,r}$. This signals a dynamical 
phase transition in the tail of the distribution of the maximum. 
Next, we considered 
a collection of independent RBMs with initial (and resetting) positions 
uniformly distributed with a density $\rho$ over the negative half-line. 
In this case, the distribution of the global maximum up to time $t$ 
depends on how one carries out the average over the initial positions. 
In the annealed case the stochastic trajectories and the initial 
positions are averaged simultaneously, while in the quenched case, one 
averages only over the stochastic trajectories for a fixed initial 
condition, which is a typical one, i.e., the initial condition that has 
the largest probability of occurrence. In the annealed case, we showed 
that the PDF of the maximum, appropriately centered around 
$\sqrt{D/r}\,\ln (r\,t)$ and scaled by a constant factor, approaches at 
late times to a new limiting distribution characterized by a scaling 
function $G'_{\rm an}(z)$, which is different from the well known Gumbel 
law for a single particle. Thus, even though the particles are 
independent, the averaging over the initial positions modifies the 
typical distribution of the maximum. However, the large-deviation 
behavior of $M_t$ when $M \sim O(t)$ remains the same as the single 
particle case. Hence, in the annealed case, the fluctuations of the 
initial positions modify the typical behavior of the maximum, but not 
its large-deviation tails. In contrast, for the quenched case, we showed 
that the PDF of the maximum, centered at $\sqrt{D/r}\,\ln (r\,t)$ and 
scaled by a constant, does approach the limiting Gumbel form, but with 
nontrivial large-deviation tails on both sides of the typical value, 
i.e., both when $M \sim O(1)$ and $M \sim O(t)$. Thus in the quenched 
case, the fluctuations of the initial positions affect only the 
large-deviation behavior of $M_t$. Our analytical predictions, both for the 
typical as well as for the large-deviation regime of $M_t$, have been 
verified numerically to a very high precision, down to $10^{-250}$ for 
the PDF of $M_t$.

There are many directions in which this work can be extended. Here we 
considered independent RBMs distributed initially over the semi-infinite 
space. Recently, another model of resetting Brownian motions has been 
introduced where the position of the walkers are simultaneously reset to 
their initial positions but they evolve independently between two 
resetting events~\cite{Biroli_PRL,Biroli_long}. Simultaneous resetting 
makes the positions of the particles correlated and it would be 
interesting to study the distribution of $M_t$ in this correlated case 
with the semi-inifinite initial conditions as studied in this paper for 
independent RBMs. Another interesting extension would be to study the 
distribution of the extremes (both the maximum as well as the minimum) 
for a set of RBMs in higher dimensions with initial positions 
distributed uniformly over a given region of space.

\acknowledgements

SNM acknowledges the Gay-Lussac Humboldt Research Award (2019), 
awarded by the Alexander von Humboldt
foundation, that enabled a visit to the University of Oldenburg 
where this work was partially done. \Blue{SNM and GS acknowledge support from ANR Grant No. ANR- 23-CE30-0020-01 EDIPS.}
The simulations were partially performed at the HPC cluster CARL, located at the
University of Oldenburg (Germany) and funded by the DFG through its Major
Research Instrumentation Program (INST 184/157-1 FUGG) and the Ministry of
Science and Culture (MWK) of the Lower Saxony State.

%\tableofcontents

\end{document}